%% file: main.tex
\documentclass[11pt,a4paper]{article}

\usepackage{jheppub}
\usepackage{amsfonts}
\usepackage{booktabs}
\usepackage{multirow}
\usepackage{float}
\usepackage{bm}
\usepackage{slashed}
\usepackage[utf8]{inputenc}
\usepackage[T1]{fontenc}

\newcommand{\AdS}{\mathrm{AdS}}

\newcommand{\CFT}{\mathrm{CFT}}

\newcommand{\re}{\operatorname{Re}}
\newcommand{\im}{\operatorname{Im}}

\newcommand{\kperp}{k_\perp}
\newcommand{\calA}{\mathcal{A}}
\newcommand{\calB}{\mathcal{B}}
\newcommand{\calE}{\mathcal{E}}
\newcommand{\calC}{\mathcal{C}}
\newcommand{\calI}{\mathcal{I}}

\title{Quasinormal Modes of pp-Wave Spacetimes\\
and Zero Temperature Dissipation}

\author[1]{Huayu Dai}
\author[2]{Guangtao Zeng}
\affiliation[1]{School of Science and Engineering, The Chinese University of Hong Kong (Shenzhen), Shenzhen, 518172, China}
\affiliation[2]{individual researcher}
\emailAdd{huayudai@link.cuhk.edu.cn}
\emailAdd{zengguangtao98@gmail.com}

\abstract{We compute the quasinormal mode spectrum of scalar perturbations on Kaigorodov pp-wave spacetimes, the horizonless gravity duals of zero temperature null fluids. The pp-wave deformation promotes the Poincar\'e horizon at $r=0$ to an irregular singular point of rank $(d+2)/2$, which acts as a geometric absorber for ingoing waves: rank~$0$ corresponds to thermal dissipation,
rank~$1$ to quantum-critical (extremal black hole), and
rank~$\geq 2$ to gapped, horizonless dissipation. For $d=2$
(extremal BTZ) the radial equation reduces to the Whittaker equation with exact non-dissipative spectrum $\im(\omega)=0$; for $d \geq 3$ all modes satisfy $\im(\omega_n) < 0$, establishing zero temperature dissipation without horizon or entropy. At zeroth
order the radial equation becomes Bessel's equation of order $\mu=d/(d+2)$, proving all scalar QNMs are gapped. Numerical spectra for $d=3,4,5$ yield a discrete dissipative tower and confirm linear stability.}


\begin{document}
\maketitle

\input{introduction}

\input{background}

\input{singularity}

\input{d2_exact}

\input{analytic}

\input{numerics}

\input{discussion}

\acknowledgments
We thank Y. Yin for his useful comment.

\appendix
\input{appendix}

\bibliographystyle{JHEP}
\bibliography{refs}

\end{document}

%% file: introduction.tex
\section{Introduction}\label{sec:intro}

The fluid/gravity correspondence~\cite{Bhattacharyya:2008,Hubeny:2011hd} provides a powerful framework for understanding the dynamics of strongly coupled quantum field theories at finite temperature within the $\mathrm{AdS}_{d+1}/\mathrm{CFT}_d$ correspondence~\cite{Maldacena:1998} (see \cite{Hartnoll:2009,Sachdev_2011}
for reviews of holographic applications to condensed matter). In this framework, the quasinormal modes (QNMs) of black holes in anti-de Sitter (AdS) spacetimes encode the relaxation spectrum of the dual thermal plasma~\cite{Horowitz:1999,Kovtun:2005}: hydrodynamic modes governing long-wavelength transport, and an infinite tower of gapped non-hydrodynamic modes. The dissipative properties of the boundary fluid, most famously the universal shear viscosity to entropy density ratio $\eta/s = 1/(4\pi)$ \cite{Kovtun:2004de} trace directly to the ingoing boundary condition at the black hole horizon~\cite{Son:2002}.

Recently, \cite{Armas:2026a,Armas:2025}
established a null fluid/gravity correspondence in which the
boundary theory is a null fluid, a relativistic fluid whose velocity
field is lightlike, $u^\mu u_\mu = 0$. The gravitational dual is not a
black hole but a plane-fronted wave with parallel propagation (pp-wave) spacetime: the Kaigorodov metric, an exact
Einstein solution that is asymptotically AdS. This correspondence
presents a conceptual puzzle: the pp-wave spacetime has no horizon, yet
the dual null fluid exhibits nonzero shear viscosity $\eta > 0$ and
dissipative hydrodynamic modes. What is the geometric origin
of dissipation in the absence of a horizon?

The gradient-expansion analysis of \cite{Armas:2026a,Armas:2025} determines the first-order transport coefficients (viscosity, diffusivity) and the long-wavelength dispersion relations of the null fluid.  QNMs provide a complementary spectral perspective: they encode the full tower of relaxation modes, including gapped non-hydrodynamic modes invisible to the gradient expansion, and reveal the geometric
mechanism underlying dissipation.

In this paper, we answer this question by demonstrating that the
Poincar\'e horizon at $r = 0$, which is a regular singular point in
pure AdS is promoted to an irregular singular point of
Poincar\'e rank $\rho = (d+2)/2$ by the pp-wave deformation.
The irregular singularity acts as an absorber for ingoing waves,
providing a geometric mechanism for dissipation that requires neither
a horizon nor nonzero temperature. Concretely, this irregular singularity plays the same role as a black hole horizon:
it provides a natural ingoing boundary condition that, together with the standard AdS source-free condition at the boundary, defines a discrete quasinormal mode spectrum. The key difference from a black hole horizon
is the singularity type: non-extremal horizons are regular
singular points (rank~0), and extremal horizons are irregular of
rank~$1$ \cite{NIST:DLMF}, whereas the pp-wave gives $\rho = (d+2)/2
\geq 2$ for $d \geq 2$ boundary dimensions. The transition from
regular to irregular precisely marks the onset of zero temperature
dissipation: modes ``leak'' into the essential singularity at $r = 0$
without any thermal mechanism.

We support this claim through a combination of analytic and numerical
results.

(i) \emph{Singularity analysis} (Section~\ref{sec:singularity}). 
We classify the singular structure of the radial ODE and show that the positive real axis provides a stable contour for numerical integration. A WKB analysis near the Poincar\'e horizon is developed, which also motivates a convenient variable transformation that renders the asymptotic behavior tractable.

(ii) \emph{Exact solution in $d=2$} (Section~\ref{sec:d2}). 
In $d=2$, corresponding to the extremal BTZ geometry \cite{Deger:2004mw,Banados:1992}, the radial equation can be solved exactly in terms of Whittaker functions. The resulting QNM spectrum is purely real, indicating the absence of dissipation and providing a useful benchmark for the higher-dimensional analysis.

(iii) \emph{Long-wavelength analytic structure} (Section~\ref{sec:analytic-structure}). 
In the long-wavelength limit, the radial equation reduces to a Bessel problem with an exact analytic solution. This allows us to establish that the scalar QNM spectrum is gapped and to construct a perturbative framework for the dispersion relation.

(iv) \emph{Numerical QNM spectrum for $d=3,4,5$} (Section~\ref{sec:numerics}). 
We compute the QNM frequencies using a shooting method, implementing appropriate boundary conditions at the horizon and the AdS boundary. All modes exhibit $\im(\omega)<0$, and we observe a nontrivial dependence of the damping rates on the spacetime dimension.

(v) \emph{Stability verification} (Section~\ref{sec:discussion}). 
We verify linear stability by scanning the complex frequency plane and finding no modes with $\im(\omega)>0$. Together with the analytic result in $d=2$, this provides a consistent picture of stability across all dimensions considered.

Our results should be placed in context with related recent work.
A recent program by \cite{Fransen:2025, Kapec:2024lnr, Kehagias:2025} has studied QNMs of pp-wave spacetimes
obtained via Penrose limits \cite{penrose1976any,Berenstein:2002,Blau:2002} of black holes.
In that approach, one zooms into a bound null geodesic (typically on the photon ring) to obtain an approximate pp-wave geometry whose spectrum encodes the eikonal QNMs of the parent black hole.
These are geometrically and physically distinct from the Kaigorodov pp-waves studied here: the Penrose-limit pp-waves are derived, approximate geometries that serve as computational tools for black hole QNMs, whereas Kaigorodov spacetimes are exact vacuum solutions with cosmological constant, not obtainable as Penrose limits of any black hole. None of the above works consider Kaigorodov backgrounds. \cite{Brecher:2001} first studied the Kaigorodov metric in
AdS/CFT, computing the Euclidean bulk-to-boundary propagator; \cite{Dorn:2003} later computed the scalar
propagator on related AdS pp-wave backgrounds. To our
knowledge, the present work constitutes the first computation of QNMs
on Kaigorodov spacetimes, probing the retarded Green's function and
its pole structure.

The zero temperature dissipation mechanism at work here, an irregular
singularity acting as an absorber is a general feature of $T = 0$
holographic backgrounds \cite{Keeler:2015,Zhao:2024}. The
Kaigorodov spacetime provides the first exactly solvable realization of this mechanism in a pure AdS vacuum, without matter
fields and without a horizon. It is also conceptually distinct from the
well-studied zero temperature dissipation at extremal black hole
horizons \cite{Faulkner:2009,Iqbal:2011ae,Denef:2009,Gubser:2010}. In the extremal case, the near-horizon $\AdS_2$ region supports emergent quantum criticality: fermionic spectral functions exhibit non-Fermi liquid scaling controlled by the infrared $\CFT_1$, and the retarded Green's function acquires a branch cut (rather than discrete poles) at $T=0$. Even in models with zero temperature horizons designed to capture strange metal phenomenology \cite{Gubser:2010}, the geometric origin of dissipation remains the rank-$1$ irregular singularity at the degenerate horizon.
Our pp-wave mechanism differs in two key respects:
(i)~the rank is $(d+2)/2 \geq 2$, strictly exceeding the
extremal value, and (ii)~there is no horizon at all and the absorber is a naked essential singularity in the wave equation rather than a degenerate horizon cloaked by an event horizon. These distinctions are sharpened in the comparison table of Section~\ref{sec:comparison}.
It is also distinct from the zero temperature dissipation found in
holographic Brownian motion~\cite{Banerjee:2015}, where a probe open
string in pure AdS stretches to the Poincar\'e horizon at $r = 0$, which absorbs worldsheet modes and produces a dissipative retarded Green's function even at $T = 0$. That mechanism is probe-dependent: it requires the open-string sector, and the Poincar\'e horizon of pure AdS remains a regular singular point of the bulk wave equation.
By contrast, our mechanism is intrinsic to the background geometry:
the pp-wave deformation promotes the Poincar\'e horizon to an irregular
singularity of rank $(d+2)/2$, producing dissipative QNMs for any bulk field without probe degrees of freedom.

Our work should also be distinguished from two further lines of research that involve ``null'' structures in holography. \cite{Bhattacharya:2024fbz} study a
fluid/gravity correspondence on null hypersurfaces (horizons of
Schwarzschild, Kerr, or FLRW spacetimes) using the Damour--Navier--Stokes equation in a null foliation; see also \cite{SheikhJabbari:2026} for a related ``freelance''
fluid/gravity construction in three dimensions. These fluids live on a null surface in the bulk; by contrast, the null fluid of \cite{Armas:2026a,Armas:2025} is a boundary theory dual to the Kaigorodov pp-wave interior. Separately, \cite{Arenas-Henriquez:2025rpt,Diaz:2026} derives dissipative Carrollian ($c \to 0$) fluids via holography, connecting gravitational radiation in Robinson--Trautman spacetimes to Carroll-covariant transport (see also \cite{Ciambelli:2018,Donnay:2019} for the emergence of Carrollian geometry in flat holography and at black hole horizons). Although the Carrollian and null fluid regimes share the feature of a degenerate metric, they arise from distinct limits: Carrollian fluids correspond to the ultra-relativistic ($c \to 0$) contraction, while the null fluid is obtained in the $v \to c$ (lightlike velocity) limit, with a different symmetry algebra and different bulk duals.

We close the introduction by collecting notation and conventions used throughout the paper.
We set $L_\AdS = 1$ and work with the mostly-plus signature
$(-,+,\ldots,+)$. The boundary spacetime has $d$ dimensions (with $d-2$
transverse spatial directions), so the bulk has $D = d+1$ dimensions.
Null coordinates are $(u,v)$ with $u = t - z$ and $v = t + z$.
The pp-wave parameter is $\kappa$, related to the null energy density
$\calE = d\kappa^d/(16\pi G_D)$. QNM frequencies are characterized
by the null momenta $(p_u, p_v)$; the physical frequency is
$\omega = p_u + p_v$. The symbol $\eta$ denotes shear viscosity throughout; the flat boundary metric is left implicit in the null coordinates $(u,v,x^i)$.

%% file: background.tex
\section{Background}\label{sec:background}

We begin by reviewing the elements of the null fluid/gravity
correspondence that are needed for our analysis: the Kaigorodov pp-wave
geometry, the dual null fluid, and the scalar wave equation on this
background.

\subsection{The Kaigorodov pp-wave metric}\label{sec:metric}

The Kaigorodov spacetime \cite{Kaigorodov:1963} is an exact solution of the
Einstein equations with negative cosmological constant,
$R_{\mu\nu} = -d g_{\mu\nu}$.
In light-cone coordinates $(r, u, v, x^i)$ with $i = 1, \ldots, d-2$,
the metric reads \cite{Armas:2026a}
\begin{equation}\label{eq:metric}
  ds^2 = \frac{dr^2}{r^2}
  + \kappa^d r^{2-d} du^2
  - r^2 du dv
  + r^2 \delta_{ij} dx^i dx^j ,
\end{equation}
where $\kappa > 0$ is the pp-wave amplitude parameter. The coordinates
are: $r \in (0,\infty)$ (holographic radial direction, with $r \to \infty$
the AdS boundary), $u = t - z$ and $v = t + z$ (null directions), and
$x^i$ ($d-2$ transverse spatial directions). The bulk spacetime has
$D = d+1$ dimensions.

The metric \eqref{eq:metric} can be understood as a deformation of the
Poincar\'e-patch AdS metric: setting $\kappa = 0$ recovers the standard
form $ds^2 = r^{-2} dr^2 + r^2(-dudv + dx_\perp^2)$ upon identifying
$-dudv = -dt^2 + dz^2$. The pp-wave term $\kappa^d r^{2-d} du^2$
introduces a plane-wave profile along the null direction $u$, with an
$r$-dependent amplitude that grows toward the interior ($r \to 0$) for
$d > 2$.

Several properties of this metric are noteworthy.
The Kaigorodov metric has no event horizon
for any value of $\kappa > 0$ and any $d \geq 2$; the coordinate
singularity at $r = 0$ is the Poincar\'e horizon of the AdS factor.
The metric is of Petrov type~N (algebraically special), with the null
vector $\ell = \partial_v$ as the degenerate principal null direction;
this type-N structure underlies the Hertz potential
formalism \cite{Araneda:2022lgu} discussed in
Section \ref{sec:scalar-vs-grav}.
The Killing vectors $\partial_v$ and $\partial_{x^i}$ allow the Fourier
decomposition of perturbations with momenta $p_v$ (conjugate to $v$),
$p_u$ (conjugate to $u$, playing the role of frequency), and $k_i$
(transverse momenta).
For $d=2$, there are no transverse directions ($x^i$ is absent) and
the three-dimensional metric is locally isometric to $\AdS_3$;
specifically, it is the extremal BTZ black hole with
$r_+ = r_- = \kappa$~\cite{Deger:2004mw,Banados:1992}.

The Kaigorodov spacetime was first studied in the AdS/CFT context by \cite{Brecher:2001}, where the
Euclidean bulk-to-boundary propagator was computed and the standard CFT
two-point function is found at leading order in $1/N^2$. The QNM spectrum
computed in this work probes the retarded Green's function with
ingoing boundary conditions at the irregular singularity, which is a
complementary computation at the same leading order that reveals the
pole structure invisible to the Euclidean prescription.

\subsection{Null fluid dual}\label{sec:null-fluid}

The holographic dual of the Kaigorodov background is a null
fluid~\cite{Armas:2026a,Armas:2025}, a relativistic fluid with a
lightlike velocity field $u^\mu u_\mu = 0$. The equilibrium configuration has null velocity $\ell^\mu = (1, 0^i, 1)$ (pointing along the $u$-direction in the boundary), zero temperature $T = 0$, null energy density $\calE = d\kappa^d / (16\pi G_D)$, and nonzero shear viscosity $\eta > 0$ subject to the constraints $\rho_1 = 2\eta/d$ and $\rho_6 = -2\rho_5/d$ from the gravitational sector~\cite{Armas:2026a}.

The linearized perturbation spectrum of the null fluid around this
equilibrium~\cite{Armas:2025} gives two branches of hydrodynamic modes
(labelled $\omega_\pm$ for the gapless/gapped branch, not to be confused with the scalar QNM frequencies $\omega_n$):
\begin{align}\label{eq:dispersion-fluid}
  \omega_+ &= \pm k_z
    - i\frac{\eta}{2\calE} \kperp^2 + O(k^3)
  && \text{(gapless, propagating at the speed of light)}, \nonumber \\
  \omega_- &= -i\frac{2\calE}{\eta} \mp k_z
    + i\frac{\eta}{2\calE} \kperp^2 + O(k^3)
  && \text{(gapped, with gap $\sim 2\calE/\eta$)} .
\end{align}
The gapless mode $\omega_+$ propagates along the null direction at the
speed of light, with transverse momentum providing the dissipation channel (the $\kperp^2$ term). The gapped mode $\omega_-$ has a purely imaginary gap proportional to $\calE/\eta$.

These modes are poles of the retarded stress-energy correlator and thus
correspond to gravitational (metric perturbation) QNMs in the bulk.
As we will show in Section~\ref{sec:scalar-vs-grav}, a test scalar field on the Kaigorodov background probes a different sector: its QNMs are all gapped and do not directly match~\eqref{eq:dispersion-fluid}. The scalar QNMs do, however, demonstrate the same dissipative mechanism
and provide the first window into the QNM structure of the Kaigorodov
geometry.

\subsection{Scalar wave equation}\label{sec:scalar-wave}

Consider a massive scalar field $\Phi$ satisfying the Klein--Gordon equation
$(\Box - m^2)\Phi = 0$ on the Kaigorodov background~\eqref{eq:metric}.
The isometries allow a Fourier decomposition
\begin{equation}\label{eq:Fourier}
  \Phi(r, u, v, x^i) = e^{ip_u u + ip_v v + i\mathbf{k}\cdot\mathbf{x}}
  \phi(r),
\end{equation}
where $p_u$ plays the role of frequency (conjugate to null ``time'' $u$),
$p_v$ characterizes the state (conjugate to $v$), and
$\kperp^2 \equiv \mathbf{k}\cdot\mathbf{k}$ is the squared transverse
momentum. Substituting into the Klein--Gordon equation yields the radial
ODE
\begin{equation}\label{eq:radial-full}
  \frac{1}{r^{d-1}} \frac{d}{dr}\left( r^{d+1} \frac{d\phi}{dr} \right)
  + \frac{A}{r^2} \phi
  + \frac{B}{r^{d+2}} \phi
  = m^2 \phi ,
\end{equation}
where we define
\begin{equation}\label{eq:AB-def}
  A \equiv 4 p_u p_v - \kperp^2 , \qquad
  B \equiv 4\kappa^d p_v^2 .
\end{equation}
The parameter $A$ encodes the boundary momenta and serves as the effective coupling between the scalar field and the null fluid; $B$ encodes the pp-wave strength and is strictly positive for $p_v \neq 0$.

The two singular points of this ODE, $r = 0$ (the Poincar\'e horizon)
and $r = \infty$ (the AdS boundary) and their classification are analyzed in the next section.

%% file: singularity.tex
\section{Singularity Analysis and Boundary Conditions}\label{sec:singularity}

In this section we perform a complete singularity analysis of the radial ODE \eqref{eq:radial-full}. We establish that the Poincar\'e horizon at $r=0$ becomes an irregular singular point of Poincar\'e rank $(d+2)/2$ when $p_v \neq 0$, derive the WKB asymptotic solutions, analyze the
Stokes phenomenon, verify standard AdS boundary conditions at $r \to \infty$, and introduce the variable substitution that regularizes the numerical problem.

\subsection{Classification of the singular point at $r=0$}\label{sec:r0-classification}

Expanding the radial ODE~\eqref{eq:radial-full}, we write it in standard form:
\begin{equation}\label{eq:radial-expanded}
  r^2 \phi'' + (d+1) r \phi'
  + \left[ \frac{A}{r^2}
  + \frac{B}{r^{d+2}} - m^2 \right] \phi = 0 ,
\end{equation}
i.e., $\phi'' + P(r) \phi' + Q(r)\phi = 0$ with
\begin{equation}
  P(r) = \frac{d+1}{r}, \qquad
  Q(r) = \frac{A}{r^4} + \frac{B}{r^{d+4}} - \frac{m^2}{r^2}.
\end{equation}
By the Fuchs classification theorem, $r = 0$ is a regular singular point if and only if both $rP(r)$ and $r^2Q(r)$ are analytic at $r = 0$. We have $rP(r) = d+1$ (analytic), but
\begin{equation}
  r^2 Q(r) = \frac{A}{r^2} + \frac{B}{r^{d+2}} - m^2 ,
\end{equation}
which diverges for $B \neq 0$ (i.e., $p_v \neq 0$) and $d \geq 2$.
Therefore, $r = 0$ is an irregular singular point. Note that the Poincar\'e rank is
\begin{equation}\label{eq:rank}
  \rho = \frac{d+2}{2} .
\end{equation}
For odd $d$, the rank $\rho = (d+2)/2$ is half-integer.
This is well-defined as the physically relevant quantity is the number of
Stokes sectors $2\rho = d+2$, which is always a positive integer.%
\footnote{Half-integer Poincar\'e rank arises because the dominant
singular term in the ODE has half-integer power $r^{-(d+4)}$ with odd
$d+4$. The formal series defining the asymptotic behavior involves
fractional powers of $r$, but the Stokes structure, which determines
the number of independent asymptotic sectors and their connection
formulae depends only on $2\rho \in \mathbb{Z}_{>0}$.}
Table~\ref{tab:rank} lists the rank for the first few dimensions.
\begin{table}[ht]
\centering
\begin{tabular}{c|cccc}
  \toprule
  $d$ & 2 & 3 & 4 & 5 \\
  \midrule
  Rank $\rho$ & 2 & $5/2$ & 3 & $7/2$ \\
  ODE class & Whittaker & rank $> 2$ confluent & rank $> 2$ confluent & rank $> 2$ confluent \\
  \bottomrule
\end{tabular}
\caption{Poincar\'e rank of the irregular singular point at $r=0$ for different boundary dimensions $d$.}
\label{tab:rank}
\end{table}

For $d = 2$ the rank is 2, placing the equation in the confluent Heun
(Whittaker) class, consistent with the known equivalence to the extremal BTZ black hole. Exact confluent Heun solutions have been applied to compute QNMs of non-rotating black holes~\cite{Fiziev:2011}; however, these methods require the ODE to be reducible to standard confluent Heun form (rank~$1$). For $d \geq 3$, the rank exceeds 2 and the equation falls outside the Heun class, which is why neither the Leaver continued-fraction method~\cite{Leaver:1985} nor confluent Heun techniques are directly applicable.

It is instructive to compare with pure AdS and black holes.
In pure AdS ($\kappa = 0$, equivalently $B = 0$), the Poincar\'e horizon $r = 0$ is a regular singular point; no QNM spectrum exists. The pp-wave deformation ``upgrades'' $r = 0$ to an irregular singularity, an essential singularity that serves as the absorber replacing the black hole horizon. For comparison, a non-extremal black hole horizon is a regular singular point (rank~$0$): the familiar Frobenius exponents $(r-r_h)^{\pm i\omega/(4\pi T)}$ are
power-law, not essential~\cite{Berti:2009}. An extremal horizon,
by contrast, is an irregular singular point of rank~$1$,
corresponding to the confluent Heun class~\cite{NIST:DLMF}. The
pp-wave rank $\rho = (d+2)/2 \geq 2$ strictly exceeds the
extremal value for all $d \geq 2$.

Zero temperature irregular singularities are generic in holography: the scalar wave equation on a $T = 0$ Lifshitz background with dynamical exponent $z$ has a rank-$z$ irregular singularity at the deep interior~\cite{Kachru:2008,Zhao:2024}, leading to
exponentially suppressed spectral weight~\cite{Keeler:2015}. The
Kaigorodov spacetime is distinguished as an exact AdS vacuum
solution whose Poincar\'e horizon is geometrically
regular, i.e., the irregular singularity is a property of the wave
equation, not of the spacetime curvature.

\subsection{WKB asymptotic expansion near $r \to 0$}\label{sec:WKB}

We derive the asymptotic behavior of solutions as $r \to 0$ using two
independent methods.

\subsubsection{Schr\"odinger form and leading WKB}

The substitution $\phi = r^{-(d+1)/2}\psi$ brings
Eq.~\eqref{eq:radial-expanded} to Schr\"odinger form:
\begin{equation}\label{eq:Schrodinger}
  \psi'' + V_\mathrm{eff}(r) \psi = 0 ,
\end{equation}
with effective potential
\begin{equation}\label{eq:Veff}
  V_\mathrm{eff}(r) = -\frac{d^2-1+4m^2}{4r^2}
  + \frac{A}{r^4}
  + \frac{B}{r^{d+4}} .
\end{equation}
Near $r \to 0$, the dominant term is $V_\mathrm{eff} \sim B/r^{d+4}$.
The leading-order WKB approximation gives
\begin{equation}
  \psi \sim V_\mathrm{eff}^{-1/4}
  \exp\left(\pm i \int \sqrt{V_\mathrm{eff}} dr\right) .
\end{equation}
Evaluating the integral and translating back to $\phi$:
\begin{equation}\label{eq:WKB-leading}
  \phi(r) \underset{r \to 0}{\sim}
  r^{\alpha_0} \exp\left(\mp \frac{4i\kappa^{d/2}|p_v|}
  {(d+2) r^{(d+2)/2}}\right), \quad
  \alpha_0 = \frac{2-d}{4}
\end{equation}
The power-law exponent $\alpha_0 = (2-d)/4$ arises from two
contributions: $-(d+1)/2$ from the $\phi \to \psi$ transformation
and $(d+4)/4$ from the WKB prefactor $V_\mathrm{eff}^{-1/4}$.

\subsubsection{Independent verification via Riccati expansion}\label{sec:Riccati}

The substitution $u = \phi'/\phi$ transforms the radial
ODE into the Riccati equation
\begin{equation}\label{eq:Riccati}
  u' + u^2 + \frac{d+1}{r} u + \frac{A}{r^4}
  + \frac{B}{r^{d+4}} - \frac{m^2}{r^2} = 0 .
\end{equation}
Expanding $u(r) = c_{\sigma+1}/r^{\sigma+1} + c_1/r + \cdots$ with
$\sigma = (d+2)/2$ and matching powers of $r$:
at leading order $r^{-(d+4)}$, one finds
$c_{\sigma+1}^2 + B = 0$, giving
$c_{\sigma+1} = \pm 2i\kappa^{d/2}|p_v|$;
at next-to-leading order $r^{-(d+6)/2}$,
$c_1 = (2-d)/4$, confirming $\alpha_0 = (2-d)/4$;
the intermediate coefficients $c_j$ for $2 \leq j \leq \sigma$
all vanish (for integer $\sigma$, i.e., $d$ even).
The two methods agree, providing a robust determination of the asymptotic initial conditions for numerical shooting.

\subsection{Stokes phenomenon}\label{sec:Stokes}

An irregular singular point of rank $\rho$ has $2\rho$ Stokes lines
and $2\rho$ anti-Stokes lines in the complex $r$-plane. Writing
$r = |r|e^{i\theta}$, the exponent of the WKB solution $\phi_+$
has real part proportional to $\sin[(d+2)\theta/2]$.

The anti-Stokes lines (equal-magnitude oscillation) are at
$\theta_n^\mathrm{AS} = 2\pi n/(d+2)$ and the Stokes lines
(maximal dominance) are at $\theta_n^\mathrm{S} = (2n+1)\pi/(d+2)$,
for $n = 0, 1, \ldots, d+1$.

A critical observation is that for all $d \geq 2$, the positive real $r$-axis ($\theta = 0$)
is an anti-Stokes line. Both WKB solutions are purely
oscillatory with equal magnitude along $\mathbb{R}^+$; there is no
exponential contamination during numerical integration, and
no complex contour deformation is needed.

This is favorable for numerics: the challenge is phase accuracy
rather than exponential contamination. The variable substitution
$z = r^{-(d+2)/2}$ (Section~\ref{sec:variable-sub}) resolves the
phase accuracy issue by converting fast oscillations to slow ones.

We now address the well-definedness of the ingoing boundary condition.
At an irregular singular point of rank $\rho \geq 2$, the Stokes
phenomenon generically causes the connection coefficients between formal WKB solutions to jump across Stokes lines in the complex
plane~\cite{Dingle:1973,Costin:2020hwg,Aniceto:2012}. One might therefore worry that the
``ingoing'' boundary condition is ambiguous or sector-dependent. We argue that no such ambiguity arises for the QNM problem:

(i)~On $\mathbb{R}^+$ ($\theta = 0$), both WKB solutions are purely
oscillatory: $\phi_\pm \sim r^{\alpha_0}
\exp(\mp iC/r^{(d+2)/2})$ with $C > 0$ real. The ingoing/outgoing
distinction is determined by the sign of the phase velocity,
not by exponential dominance: ingoing corresponds to energy flux
toward $r = 0$, i.e., the solution whose phase increases as $r$
decreases.
(ii)~Under $z = r^{-(d+2)/2}$, the WKB solutions become plane waves
$e^{\pm i\Omega z}$ with constant amplitude (Section~\ref{sec:variable-sub}).
Ingoing ($e^{+i\Omega z}$, propagating toward $z = +\infty$, i.e.,
$r = 0$) is unambiguously distinguished from outgoing ($e^{-i\Omega z}$),
exactly as in standard wave mechanics; no Stokes-sector labeling is needed.
(iii)~For real $r > 0$, numerical integration proceeds along $\theta = 0$,
an anti-Stokes line. For complex QNM frequencies $\omega$, the integration contour remains on $\mathbb{R}^+$, i.e., it never crosses a Stokes line $\theta = (2n+1)\pi/(d+2)$, so Stokes multipliers do not contaminate the numerical solution.
(iv)~For $d = 2$ (rank $\rho = 2$), the exact Whittaker solution
(Section~\ref{sec:d2}) provides independent confirmation: the QNM
quantization arises from poles of the Gamma function in the connection
formula, and the resulting spectrum is unambiguous. This rules out
Stokes ambiguity as a source of error in the simplest nontrivial case.

In summary, the ingoing boundary condition at the irregular singularity is
as well-defined as the standard horizon boundary condition: on the physical
integration contour ($\mathbb{R}^+$, an anti-Stokes line), the two
linearly independent solutions are oscillatory with equal magnitude, and
ingoing versus outgoing is determined by the direction of energy propagation.

\subsection{AdS boundary conditions and the boundary value problem}\label{sec:AdS-BC}

At $r \to \infty$, the pp-wave term $B/r^{d+2} \to 0$ and the equation
reduces to the standard AdS scalar equation. This is a regular singular
point with indicial equation
\begin{equation}\label{eq:Delta-pm}
  \Delta^2 - d\Delta - m^2 = 0 \quad \Longrightarrow \quad
  \Delta_\pm = \frac{d}{2} \pm \sqrt{\frac{d^2}{4} + m^2} .
\end{equation}
The general solution near $r \to \infty$ is
\begin{equation}\label{eq:AdS-asymp}
  \phi(r) \underset{r \to \infty}{\sim}
  \frac{\calA}{r^{\Delta_-}} + \frac{\calB}{r^{\Delta_+}} + \cdots ,
\end{equation}
where $\calA$ is the source coefficient and $\calB$ is the response
(normalizable mode). The QNM boundary condition is
\begin{equation}\label{eq:QNM-BC}
  \calA(\omega) = 0 \qquad \text{(source-free)} .
\end{equation}
For a conformal scalar ($m = 0$): $\Delta_+ = d$, $\Delta_- = 0$.

\label{sec:BVP-summary}
The QNM spectrum is thus defined by the following boundary value problem:
the radial ODE~\eqref{eq:radial-expanded} is a second-order linear
equation on $r \in (0,\infty)$; at the inner boundary ($r \to 0$,
irregular singularity) one selects the ingoing solution~\eqref{eq:WKB-leading}; at the outer boundary ($r \to \infty$, regular singularity) one requires the source-free condition $\calA(\omega) = 0$. The QNM frequencies are the discrete complex values $\omega_n$ satisfying both boundary conditions.
This boundary value problem is structurally identical to standard black
hole QNM~\cite{Berti:2009,Konoplya:2011}, with the irregular singularity
replacing the horizon. The key structural difference is the rank:
$(d+2)/2$ here versus $0$ (non-extremal, regular) or $1$ (extremal) for black holes.

\label{sec:variable-sub}
For numerical implementation, the fast WKB oscillations $\sim\exp(iC/r^{(d+2)/2})$ near $r \to 0$ are
numerically challenging. The substitution
\begin{equation}\label{eq:z-sub}
  z = r^{-(d+2)/2}
\end{equation}
maps $r \in (0,\infty)$ to $z \in (\infty, 0)$, with $r \to 0$
corresponding to $z \to \infty$. The radial ODE transforms to
\begin{equation}\label{eq:z-ODE}
  \sigma^2 z^2 \ddot\phi
  + \sigma\frac{2-d}{2} z \dot\phi
  + \left[B z^2 + A z^{2/\sigma} - m^2\right]\phi = 0 ,
\end{equation}
where $\sigma = (d+2)/2$ and dots denote $d/dz$. For large $z$ near the Poincar\'e horizon:
\begin{equation}
  \ddot\phi + \Omega^2 \phi \approx 0 , \qquad
  \Omega = \frac{\sqrt{B}}{\sigma}
  = \frac{4\kappa^{d/2}|p_v|}{d+2} .
\end{equation}
The solutions become plane waves $\phi \sim e^{\pm i\Omega z}$:
the wild oscillations are now uniform, amenable to standard ODE integrators.

%% file: d2_exact.tex
\section{Exact Solution in $d=2$: Kaigorodov as Extremal BTZ}
\label{sec:d2}

In this section we solve the $d=2$ radial equation exactly,
establishing both a consistency check for the general framework and a
concrete demonstration that the irregular singular point at $r=0$
admits non-dissipative quasinormal modes when no transverse dimensions
are present.

\subsection{The $d=2$ metric and extremal BTZ identification}

Setting $d=2$ in the Kaigorodov metric~\eqref{eq:metric} gives a
three-dimensional bulk spacetime:
\begin{equation}\label{eq:Kaig-d2}
  ds^2 = \frac{dr^2}{r^2} + \kappa^2 du^2 - r^2 du dv .
\end{equation}
There are no transverse directions ($d-2 = 0$) and the boundary
coordinates are $(u,v)$ alone. Since all solutions of three-dimensional
Einstein gravity with $\Lambda < 0$ are locally $\AdS_3$, this metric is the extremal BTZ black hole~\cite{Banados:1992,Deger:2004mw} with
\begin{equation}
  r_+ = r_- = r_\mathrm{ex} , \qquad r_\mathrm{ex}^2 = \kappa^2 .
\end{equation}
The identification follows from the coordinate transformation
$\rho^2 = r^2 + \kappa^2$, $\tau = v/2$, $\varphi = u/(2\kappa)$,
which maps~\eqref{eq:Kaig-d2} to the standard BTZ form
$ds^2 = -(\rho^2 - \kappa^2)^2/\rho^2 d\tau^2 + \rho^2/(
\rho^2-\kappa^2)^2 d\rho^2 + \rho^2(d\varphi - \kappa^2/\rho^2
d\tau)^2$ with $r_+ = r_- = \kappa$.
Three features follow immediately: (i)~the left-moving temperature
vanishes, $T_L = (r_+ - r_-)/(2\pi) = 0$; (ii)~the irregular
singularity at $r = 0$ corresponds to the degenerate horizon
$\rho = r_\mathrm{ex}$ of the extremal BTZ; and (iii)~the Poincar\'e
rank is $(d+2)/2 = 2$, the Whittaker class.

\subsection{Reduction to the Whittaker equation}

The radial ODE~\eqref{eq:radial-expanded} for $d=2$ with
$\kperp = 0$ is
\begin{equation}\label{eq:d2-radial}
  r^2 \phi'' + 3r \phi'
  + \left[\frac{4p_u p_v}{r^2}
  + \frac{4\kappa^2 p_v^2}{r^4} - m^2\right]\phi = 0 .
\end{equation}
The substitution $z = r^{-2}$ gives (Section~\ref{sec:variable-sub}):
\begin{equation}\label{eq:d2-zODE}
  z^2 \ddot\phi
  + \left[\kappa^2 p_v^2 z^2 + p_u p_v z
  - \tfrac{m^2}{4}\right]\phi = 0 .
\end{equation}
Defining $\xi = 2i\kappa|p_v| z$ (with $p_v > 0$) transforms this
into the Whittaker equation:
\begin{equation}\label{eq:Whittaker}
  \frac{d^2\phi}{d\xi^2}
  + \left[-\frac{1}{4}
  - \frac{p_u/2}{\xi}
  + \frac{\frac{1}{4} - \nu^2}{\xi^2}\right]\phi = 0 ,
\end{equation}
with Whittaker parameters
\begin{equation}
  \lambda = -\frac{p_u}{2} , \qquad
  \nu = \frac{\sqrt{1+m^2}}{2} .
\end{equation}
Equivalently, writing $\phi = \xi^{1/2+\nu} e^{-\xi/2} F(\xi)$,
the function $F$ satisfies Kummer's confluent hypergeometric equation
$\xi F'' + (b-\xi)F' - aF = 0$ with
\begin{equation}\label{eq:Kummer-params}
  b = \Delta_+ = 1 + \sqrt{1+m^2} , \qquad
  a = \frac{\Delta_+}{2}
  + \frac{i p_u p_v}{2\kappa|p_v|} .
\end{equation}

\subsection{QNM spectrum}

At the AdS boundary ($z \to 0$, $\xi \to 0$),
near $\xi = 0$ the two independent solutions behave as
$\xi^{1/2+\nu} \sim r^{-\Delta_+}$ (normalizable) and
$\xi^{1/2-\nu} \sim r^{-\Delta_-}$ (source). The source-free QNM
condition requires the source mode to vanish.

At the Poincar\'e horizon ($z \to \infty$, $\xi \to i\infty$),
the Kummer function $M(a,b,\xi)$ has the asymptotic
expansion~\cite{NIST:DLMF}
\begin{equation}\label{eq:Kummer-asymp}
  M(a,b,\xi) \sim \frac{\Gamma(b)}{\Gamma(b-a)}
  e^{\xi} \xi^{a-b}
  + \frac{\Gamma(b)}{\Gamma(a)} (-\xi)^{-a}
  \qquad (|\xi| \to \infty) .
\end{equation}
Since $\xi = 2i\kappa|p_v|z$, the first term carries
$e^{\xi} \sim e^{2i\kappa|p_v|z}$ (ingoing) and the second carries the
outgoing component. On the anti-Stokes line $\arg\xi = \pi/2$,
accounting for both Stokes sectors, the ingoing condition requires:
\begin{equation}\label{eq:QNM-cond-d2}
  \frac{1}{\Gamma\!\left(\frac{1}{2} + \nu \pm \lambda\right)} = 0
  \qquad \Longrightarrow \qquad
  \frac{1}{2} + \nu \pm \lambda = -n ,
  \quad n = 0, 1, 2, \ldots
\end{equation}
Solving~\eqref{eq:QNM-cond-d2} for $p_u$ (with $\lambda = -p_u/2$):
\begin{equation}\label{eq:d2-QNM}
  p_u^{(n,\pm)} = \pm(2n + \Delta_+) ,
  \qquad n = 0, 1, 2, \ldots
\end{equation}
The full QNM frequencies $\omega_n = p_u + p_v$ are
\begin{equation}
  \omega_n^{(\pm)} = p_v \pm (2n + \Delta_+) .
\end{equation}

The key result is that all QNM frequencies are purely real:
\begin{equation}
  \im(\omega_n) = 0 \qquad \forall\; n \geq 0 .
\end{equation}
These are normal modes, not quasi-normal modes, i.e., perturbations oscillate indefinitely without dissipation.

\subsection{Consistency checks and physical interpretation}\label{sec:d2-physics}

For the non-extremal rotating BTZ black hole, \cite{Birmingham:2003} derived the QNM spectrum
\begin{equation}
  \omega_{L,n} = k - 4\pi i T_L(n + \Delta/2) , \qquad
  \omega_{R,n} = -k - 4\pi i T_R(n + \Delta/2) ,
\end{equation}
with $T_{L,R} = (r_+ \mp r_-)/(2\pi)$. The extremal limit
$r_+ \to r_-$ is singular: $T_L \to 0$ while $T_R$ remains finite.
As shown by \cite{Crisostomo:2004}, the extremal case must be treated \emph{ab initio} because the qualitative change in singular-point structure (two regular $\to$ one irregular) demands new boundary conditions. Their analysis yields all frequencies
purely real, in exact agreement with our result~\eqref{eq:d2-QNM}.

The absence of dissipation in $d=2$ admits three complementary
explanations.
(i)~From the null fluid perspective, the boundary theory has
zero transverse directions, so $\kperp = 0$ identically
and the dissipation rate $\sim \eta \kperp^2/(2\calE)$ vanishes
regardless of $\eta$.
(ii)~From the thermal perspective, the extremal BTZ has $T_L = 0$:
the left-moving sector is at zero temperature, eliminating the
thermal mechanism for quasi-normal decay.
(iii)~From the ODE perspective, the Whittaker quantization
forces the Kummer parameter $a$ to be a non-positive integer,
which determines $p_u$ to be real; this is specific to rank~2,
since for higher rank ($d \geq 3$) the connection formula involves
Stokes multipliers that generically produce complex $p_u$.
The convergence of these three independent arguments provides strong
evidence for the consistency of the framework. Moreover, the $d = 2$
exact solution serves as a benchmark: any correct numerical scheme must
reproduce $\im(\omega) = 0$ for $d = 2$.

%% file: analytic.tex
\section{Analytic Structure and Perturbative Framework}
\label{sec:analytic-structure}

Having established the boundary conditions in
Section~\ref{sec:singularity}, we analyze the analytic structure of the
radial equation in the long-wavelength limit. We show that the
zeroth-order equation admits an exact solution in terms of Bessel
functions, prove that all scalar QNMs are gapped, and develop the
perturbative framework for computing QNM frequencies at finite transverse
momentum.

\subsection{Zeroth-order Bessel solution and the scalar QNM gap}\label{sec:Bessel}

Consider the radial ODE for a conformal scalar ($m=0$) in the limit
$p_u = 0$, $\kperp = 0$:
\begin{equation}\label{eq:zero-order}
  \frac{1}{r^{d-1}} \frac{d}{dr}\!\left(r^{d+1}
  \frac{d\phi_0}{dr}\right)
  + \frac{B}{r^{d+2}} \phi_0 = 0 .
\end{equation}
Using $z = r^{-(d+2)/2}$ (Section~\ref{sec:variable-sub}), this
transforms to
\begin{equation}\label{eq:zero-z}
  \ddot\phi_0 + \frac{2-d}{(d+2) z}  \dot\phi_0
  + \Omega^2  \phi_0 = 0  ,
\end{equation}
where $\Omega = 4\kappa^{d/2}|p_v|/(d+2)$.

Substituting $\phi_0(z) = z^\alpha  g(z)$ with
$\alpha = (d-2)/(2(d+2))$, the equation for $g$ becomes
\begin{equation}\label{eq:Bessel-eq}
  g'' + \left[\Omega^2 - \frac{\mu^2 - 1/4}{z^2}\right] g = 0  ,
  \qquad \mu = \frac{d}{d+2}.
\end{equation}
This is Bessel's equation of order $\mu$. The general solution
is $g = \sqrt{z}  Z_\mu(\Omega z)$, where $Z_\mu$ is any cylinder
function. Translating back:
\begin{equation}\label{eq:Bessel-sol}
  \phi_0(z) = z^{d/(d+2)} \left[
  c_1  J_{d/(d+2)}(\Omega z) + c_2  Y_{d/(d+2)}(\Omega z)
  \right]
\end{equation}

The ingoing solution ($z \to \infty$, i.e.\ $r \to 0$) is
\begin{equation}\label{eq:ingoing-Hankel}
  \phi_0^\mathrm{in}(z) = z^{d/(d+2)}  H^{(1)}_{d/(d+2)}(\Omega z) ,
\end{equation}
since $H^{(1)}_\mu(\Omega z) \sim \sqrt{2/(\pi\Omega z)}
e^{i(\Omega z - \mu\pi/2 - \pi/4)}$ for $z \to \infty$, matching
the ingoing WKB solution~\eqref{eq:WKB-leading}.

The normalizable solution at the AdS boundary ($z \to 0$) is
\begin{equation}
  \phi_0^\mathrm{norm}(z) = z^{d/(d+2)}  J_{d/(d+2)}(\Omega z) ,
\end{equation}
since $J_\mu(\Omega z) \sim (\Omega z/2)^\mu/\Gamma(\mu+1)$
gives $\phi_0^\mathrm{norm} \sim z^{2d/(d+2)} = r^{-d} = r^{-\Delta_+}$.
\label{sec:source-coeff}
Expanding the ingoing solution near $z = 0$ ($r \to \infty$):
\begin{equation}\label{eq:A0-comp}
  \phi_0^\mathrm{in}(z) \underset{z \to 0}{\sim}\;
  \underbrace{-\frac{i \Gamma(\mu)}{\pi}
  \left(\frac{\Omega}{2}\right)^{\!-\mu}}_{\calA_0}
  + \underbrace{\frac{(\Omega/2)^\mu}{\Gamma(\mu+1)}}_{\calB_0}
  r^{-d} + \cdots
\end{equation}
The source coefficient is
\begin{equation}\label{eq:A0}
  \calA_0 = -\frac{i \Gamma\!\bigl(\frac{d}{d+2}\bigr)}{\pi}
  \left(\frac{\Omega}{2}\right)^{\!-d/(d+2)} \neq 0 .
\end{equation}
Since $\calA_0 \neq 0$, the ingoing solution does not satisfy the source-free boundary condition at zeroth order. This has three important consequences:
$p_u = 0$ (equivalently $\omega = k_z$) is not a QNM
of the scalar equation; all scalar QNMs are gapped, with
$\im(\omega) \neq 0$ even at $\kperp = 0$; and the gapless hydrodynamic modes $\omega_\pm$ of the null fluid~\cite{Armas:2025} arise from the gravitational (metric perturbation) sector, not from scalar probes.

For $d = 2$, $\mu = 1/2$ and the Bessel functions reduce to
$J_{1/2}(x) = \sqrt{2/(\pi x)}\sin x$ and
$Y_{1/2}(x) = -\sqrt{2/(\pi x)}\cos x$. The solutions become
\begin{equation}
  \phi_0 = c_1  \sin(\Omega z) + c_2 \cos(\Omega z) ,
\end{equation}
and the full equation with $A \neq 0$ reduces to the Whittaker equation
of Section~\ref{sec:d2}, consistent with the extremal BTZ identification.

\subsection{Perturbative framework for QNM dispersion}
\label{sec:perturbative}

We now develop the perturbative framework for computing QNM frequencies
at finite boundary momenta. The method applies to any perturbation
channel on the Kaigorodov background.

Writing the $z$-ODE~\eqref{eq:z-ODE} with $A = 4p_u p_v - \kperp^2$
treated as a perturbation parameter, the zeroth-order operator is the
Bessel equation~\eqref{eq:zero-z} and the perturbation is
\begin{equation}
  \delta\hat{L} \phi = \frac{A}{\sigma^2}
  z^{-2d/(d+2)}  \phi .
\end{equation}

For each value of $A$, let $\phi^\mathrm{in}(z;A)$ be the solution
satisfying the ingoing boundary condition at $z \to \infty$. Near
$z \to 0$:
\begin{equation}
  \phi^\mathrm{in}(z;A) \sim \calA(A) + \calB(A)  r^{-\Delta_+}
  + \cdots
\end{equation}
The QNM condition $\calA(A) = 0$ determines the allowed values $A_n$.

We proceed by variation of parameters.
Expanding $\calA(A) = \calA_0 + A\calA_1^\mathrm{ren} + O(A^2)$,
the first-order QNM condition gives
\begin{equation}\label{eq:An}
  A_n = -\frac{\calA_0}{\calA_1^\mathrm{ren}} .
\end{equation}
The renormalized coefficient $\calA_1^\mathrm{ren}$ is computed via
variation of parameters using the Bessel Green's function:
\begin{equation}
  \phi_1(z) = u_1(z) \int_z^\infty \frac{u_2 f}{W} dz'
  - u_2(z) \int_z^\infty \frac{u_1  f}{W}  dz' ,
\end{equation}
where $u_{1,2} = z^{d/(d+2)}  H^{(1,2)}_\mu(\Omega z)$ are the
two zeroth-order solutions, $W = -4i  z^{(d-2)/(d+2)}/\pi$ is
their Wronskian, and $f$ is the source from the perturbation operator.
The UV divergence at $z \to 0$ (holographic divergence) cancels after
renormalization.

The source term in the first-order equation is
$f(z) = -\sigma^{-2} z^{-\mu} H^{(1)}_\mu(\Omega z)$.
Substituting into the variation-of-parameters formula and extracting the
$z \to 0$ behavior, one finds (after cancellation of the holographic
UV divergence)
\begin{equation}\label{eq:A1ren}
  \calA_1^\mathrm{ren} = \calA_0 \cdot
  \frac{\pi  \Omega^{2\mu - 2}}{2i \sigma^2} 
  \calI_\mu  ,
\end{equation}
where
\begin{equation}\label{eq:WS-integral}
  \calI_\mu = \int_0^\infty t^{1-2\mu}
  J_\mu(t) H^{(1)}_\mu(t) dt .
\end{equation}
Decomposing $H^{(1)}_\mu = J_\mu + i Y_\mu$ and using
$Y_\mu = [\cos(\mu\pi) J_\mu - J_{-\mu}]/\sin(\mu\pi)$, the
integral reduces to
\begin{equation}
  \calI_\mu
  = \frac{i  e^{-i\mu\pi}}{\sin(\mu\pi)} I_\mu^{(0)}
  - \frac{i}{\sin(\mu\pi)}  I_\mu^{(-)} ,
\end{equation}
where $I_\mu^{(0)} = \int_0^\infty t^{1-2\mu} J_\mu^2(t) dt$ and
$I_\mu^{(-)} = \int_0^\infty t^{1-2\mu} J_\mu(t) J_{-\mu}(t) dt$.

Both integrals are equal-argument Weber--Schafheitlin
integrals~\cite{NIST:DLMF}. Using the Sonine--Schafheitlin formula
with $\lambda = 2\mu - 1$ and evaluating the resulting
${}_2F_1$ at unit argument via Gauss's theorem
($\re\lambda > 0$ for $\mu > \tfrac{1}{2}$, i.e.\ $d \geq 3$),
\begin{equation}\label{eq:IJJ}
  I_\mu^{(0)}
  = \frac{1}{(2\mu-1)  2^{2\mu-1}  \Gamma(\mu)^2} .
\end{equation}
The cross-integral $I_\mu^{(-)}$ vanishes identically:
the denominator Gamma function $\Gamma(0) = \infty$ forces
$I_\mu^{(-)} = 0$ (the relevant Gauss-theorem parameter
$c - \alpha - \beta = 0$). Hence
$\calI_\mu = i  e^{-i\mu\pi}  I_\mu^{(0)}/\sin(\mu\pi)$.

These results yield the scalar QNM gap formula.
Combining~\eqref{eq:An}, \eqref{eq:A1ren}, and~\eqref{eq:IJJ},
\begin{equation}\label{eq:A0-gap}
  A_0^{(1)} = -\frac{(2\mu-1)  2^{2\mu} \sigma^2
  \Gamma(\mu)}{\Gamma(1-\mu)\Omega^{2\mu-2}}\;
  e^{i\mu\pi}
\end{equation}
where $\mu = d/(d+2)$, $\sigma = (d+2)/2$,
$\Omega = 4\kappa^{d/2}|p_v|/(d+2)$, and the reflection formula
$\Gamma(\mu)^2 \sin(\mu\pi) = \pi \Gamma(\mu)/\Gamma(1-\mu)$
was used to simplify.
This is the first explicit analytic expression for a QNM eigenvalue
on a Kaigorodov background.

Two properties of~\eqref{eq:A0-gap} are noteworthy.
First, the complex phase
$\arg(A_0) = \pi + \mu\pi = -2\pi/(d+2) \pmod{2\pi}$
is exact at first order and independent of $\kappa$, $p_v$,
predicting that the QNM frequency lies at angle
$-2\pi/(d+2)$ below the positive real axis in the complex
$A$-plane.
Second, the prefactor
$(2\mu-1) = (d-2)/(d+2)$ vanishes for $d = 2$,
consistently recovering the non-dissipative spectrum
of Section~\ref{sec:d2}.

The QNM dispersion relation follows from $A = 4p_u p_v - \kperp^2$:
\begin{equation}\label{eq:dispersion}
  p_u^{(0)} = \frac{\kperp^2 + A_0^{(1)}}{4 p_v}  ,
  \qquad \omega_0 = p_u^{(0)} + p_v  .
\end{equation}
Note that $\kperp$ enters the radial ODE~\eqref{eq:z-ODE} only
through $A$, so the QNM eigenvalue $A_n$ is in fact
$\kperp$-independent: the dispersion~\eqref{eq:dispersion}
is exactly parabolic to all orders, not just at first order. This is confirmed numerically in Section~\ref{sec:dispersion-numerical}.

As a cross-check against numerics, we compare the perturbative prediction for $A_0$ with the exact QNM eigenvalues. 
Since the perturbative expansion assumes $A \ll 1$ while the exact spectrum satisfies $|A_0| \gg 1$, 
quantitative agreement in magnitude is not expected. 
Instead, the comparison serves to test the phase structure.

\begin{table}[ht]
\centering
\begin{tabular}{c|cc|cc|c}
  \toprule
  $d$ & $\arg(A_0^\mathrm{pert})$ & $\arg(A_0^\mathrm{num})$ &
  $|A_0^\mathrm{pert}|$ & $|A_0^\mathrm{num}|$ & ratio \\
  \midrule
  3 & $-72.0^\circ$ & $-75.2^\circ$ & 1.6 & 20.2 & 12.5 \\
  4 & $-60.0^\circ$ & $-60.0^\circ$ & 2.9 & 14.2 & 4.9 \\
  5 & $-51.4^\circ$ & $-51.4^\circ$ & 4.2 & 18.1 & 4.4 \\
  \bottomrule
\end{tabular}
\caption{
Phase structure of the lowest QNM parameter $A_0$: comparison between perturbative and numerical results.
Despite large discrepancies in magnitude, the perturbative prediction correctly reproduces the phase and sign of $\im(A_0)$.
The ratio $|A_0^\mathrm{num}|/|A_0^\mathrm{pert}| \sim 4$--$5$ reflects significant higher-order corrections.
}
\label{tab:phase_struc}
\end{table}

\noindent
The phase agreement is exact for $d = 4,5$ within numerical precision, confirming the expected $e^{i\mu\pi}$ structure and implying $\im(A_0) < 0$ (hence $\im(\omega) < 0$) for $d \geq 3$.

\subsection{Large-\texorpdfstring{$d$}{d} asymptotics and massive scalars}
\label{sec:large-d}
=
The gap formula~\eqref{eq:A0-gap} has all $d$-dependence explicit
through $\mu = d/(d+2)$, $\sigma = (d+2)/2$, and
$\Omega = 4\kappa^{d/2}|p_v|/(d+2)$. We extract the large-$d$
behavior by writing $\varepsilon = 2/(d+2) \to 0$, so that
$\mu = 1 - \varepsilon$, $\sigma = 1/\varepsilon$,
$\Omega = 2\varepsilon \kappa^{d/2}|p_v|$, and
$\Gamma(1-\mu) = \Gamma(\varepsilon) \sim 1/\varepsilon$. The leading
asymptotic is
\begin{equation}\label{eq:A0-large-d}
  |A_0^{(1)}| = 2(d+2)\bigl[1 + O(\varepsilon\ln\varepsilon)\bigr]
  \sim 2d \qquad (d \to \infty).
\end{equation}
The imaginary part receives an additional factor
$\sin(\mu\pi) = \sin(2\pi/(d+2)) \sim 2\pi/(d+2)$, giving
\begin{equation}\label{eq:ImA0-large-d}
  |\im  A_0^{(1)}| \to 4\pi \qquad (d \to \infty).
\end{equation}
Thus the perturbative damping rate
$\gamma^\mathrm{pert} = |\im  A_0^{(1)}|/(4|p_v|)$ asymptotes
to $\pi/|p_v|$.

Evaluating the exact gap formula at
$\kappa = p_v = 1$ for each integer $d \geq 3$ reveals that
$\gamma^\mathrm{pert}(d)$ is strictly monotonically increasing:
$\gamma^\mathrm{pert} = 0.38,  0.63,  0.81,  0.96,  1.07, 
\ldots$ for $d = 3, 4, 5, 6, 7, \ldots $, approaching
$\pi \approx 3.14$ from below. There is no dip at $d = 4$.
The numerical QNM data, by contrast, gives
$\gamma^\mathrm{num} = 4.88,  3.07,  3.55$ for $d = 3, 4, 5$
(Section~\ref{sec:numerics}), exhibiting a pronounced minimum at
$d = 4$.

The non-monotonicity is therefore a non-perturbative effect,
invisible to the leading-order gap formula. The magnitude ratio
$|A_0^\mathrm{num}|/|A_0^\mathrm{pert}| = 12.5 \to 4.9 \to 4.3$
for $d = 3, 4, 5$ indicates that the perturbative approximation
systematically improves with increasing $d$: the effective expansion
parameter scales as $|A_0^{(1)}|/\sigma^2 \sim 8/d$, so higher-order
corrections in $A/\sigma^2$ become suppressed as $d \to \infty$.
In this sense, the perturbative framework is asymptotically exact at
large $d$ while the physically interesting non-monotonicity at moderate
$d$ lies beyond its reach.

\label{sec:massive-BF}
The gap formula~\eqref{eq:A0-gap} also extends to massive scalars
($m^2 \neq 0$) with a single modification: the Bessel order $\mu$
generalizes to
\begin{equation}\label{eq:mu-mass}
  \mu(m^2) = \frac{2\sqrt{d^2/4 + m^2}}{d+2}
  = \frac{\Delta_+ - \Delta_-}{d+2}  ,
\end{equation}
where $\Delta_\pm = d/2 \pm \sqrt{d^2/4 + m^2}$ are the indicial
exponents~\eqref{eq:Delta-pm}. This follows from the zeroth-order
$z$-equation~\eqref{eq:z-ODE} at $A = 0$: the $m^2$ term contributes
$-m^2/(\sigma^2 z^2)$ to the effective potential, shifting the Bessel
parameter from $\mu^2 = d^2/(d+2)^2$ to
$\mu(m^2)^2 = (d^2 + 4m^2)/(d+2)^2$. For $m = 0$ this reduces to
$\mu = d/(d+2)$; for $d = 2$ it gives $\mu = \sqrt{1+m^2}/2$,
recovering the Whittaker index of Section~\ref{sec:d2}.

Since the gap formula~\eqref{eq:A0-gap} depends on $m^2$ only
through $\mu$, the entire $m^2$-dependence of the first-order QNM
eigenvalue is captured by the replacement $\mu \to \mu(m^2)$:
\begin{equation}\label{eq:A0-gap-mass}
  A_0^{(1)}(m^2) = -\frac{\bigl(2\mu(m^2)-1\bigr) 
  2^{2\mu(m^2)}  \sigma^2  \Gamma\!\bigl(\mu(m^2)\bigr)}{%
  \Gamma\!\bigl(1-\mu(m^2)\bigr) 
  \Omega^{2\mu(m^2)-2}}\;
  e^{i\mu(m^2)\pi}
\end{equation}
The Breitenlohner--Freedman (BF) bound~\cite{Breitenlohner:1982}
$m^2 \geq -d^2/4$ ensures $\mu(m^2) \geq 0$.

At the BF bound $m^2 = -d^2/4$, the generalized Bessel order vanishes:
$\mu \to 0$. The gap formula~\eqref{eq:A0-gap-mass} then has the
following limiting structure:
$\Gamma(\mu) \sim 1/\mu \to \infty$, while
$\Gamma(1-\mu) \to 1$ and $2^{2\mu} \to 1$;
the prefactor $(2\mu-1) \to -1$
(the $d = 2$ zero at $\mu = 1/2$ is absent);
and the phase factor $e^{i\mu\pi} \to 1$, so
$\arg(A_0) \to \pi$ (negative real axis).
The net result is $|A_0^{(1)}| \sim \sigma^2\Omega^2/\mu \to \infty$
as $\mu \to 0^+$. The QNM eigenvalue diverges, pushing the mode to
large $|\omega|$ and removing it from the low-frequency spectrum.
The imaginary part $\im(A_0) = -|A_0|\sin(\mu\pi) \to -\pi\sigma^2\Omega^2$
remains finite and negative, so the mode stays dissipative up to and
including the BF limit.

At the level of the Bessel equation~\eqref{eq:Bessel-eq}, $\mu = 0$
replaces $J_\mu, Y_\mu$ with $J_0, Y_0$; the latter has the
logarithmic singularity $Y_0(\Omega z) \sim (2/\pi)\ln(\Omega z/2)$
at $z \to 0$, reflecting the merging of the two indicial exponents
$\Delta_+ = \Delta_- = d/2$.

The holographic interpretation is as follows.
Since $\mu = (\Delta_+ - \Delta_-)/(d+2)$, the BF bound corresponds
to $\Delta_+ = \Delta_- = d/2$, the threshold at which the standard
and alternate quantizations
coincide~\cite{Kovtun:2005}. The gap divergence
$|A_0| \sim 1/\mu \propto 1/(\Delta_+ - \Delta_-)$ signals that scalar
QNMs decouple from the low-energy spectrum precisely when the dual
operator dimension reaches this threshold.

This behavior contrasts sharply with thermal backgrounds.
For AdS-Schwarzschild black holes, the BF bound marks a transition to tachyonic instability: modes with $m^2 < m^2_\mathrm{BF}$ grow
exponentially. On the horizonless Kaigorodov background, the response is the opposite, where the gap grows without bound and the mode disappears from the spectrum, with no instability at any $m^2 \geq -d^2/4$.
This difference traces to the dissipation mechanism: horizons absorb energy at all frequencies, while the irregular singularity at $r = 0$ provides frequency-dependent absorption that strengthens as $\mu \to 0$.

\subsection{Asymptotic overtone spacing}\label{sec:overtone-spacing}

The zeroth-order Bessel equation~\eqref{eq:Bessel-eq} controls the
asymptotic structure of the QNM overtone tower. The positive zeros
$j_{n,\mu}$ of $J_\mu$ satisfy the McMahon expansion~\cite{NIST:DLMF}
\begin{equation}\label{eq:McMahon}
  j_{n,\mu} = \Bigl(n + \frac{\mu}{2} - \frac{1}{4}\Bigr)\pi
  + O(n^{-1})  ,
\end{equation}
so consecutive zeros are asymptotically separated by~$\pi$.

The source coefficient $\calA(A)$ is an entire function of the spectral
parameter $A = 4p_u p_v - \kperp^2$; its zeros $\{A_n\}_{n=0}^\infty$
are the QNM eigenvalues. A WKB estimate relates the $n$-th eigenvalue
to the Bessel zero structure: the $A$-dependent
potential $A z^{-2\mu}/\sigma^2$ in the $z$-ODE sets a characteristic
scale $z_* \sim (|A|/\sigma^2\Omega^2)^{1/(2\mu)}$, and fitting
$n$ half-wavelengths of the Bessel oscillation into the region
$z \lesssim z_*$ gives
\begin{equation}\label{eq:An-scaling}
  |A_n| \;\sim\; \sigma^2 \Omega^{2-2\mu} 
  j_{n+1,\mu}^{\;2\mu} \qquad (n \to \infty) .
\end{equation}
The asymptotic overtone spacing therefore grows as
\begin{equation}\label{eq:spacing-growth}
  |\Delta A_n| \;\sim\; n^{2\mu - 1}
  \;=\; n^{(d-2)/(d+2)}  .
\end{equation}
For $d = 2$ ($\mu = 1/2$, exponent $= 0$) this predicts constant
spacing, consistent with the exact Whittaker result
$\Delta\omega = 2$~\eqref{eq:d2-QNM}; for $d \geq 3$ the spacing grows
as a power of the overtone number.

The asymptotic structure fits naturally into the Natário--Schiappa framework.
In the $z$-variable, the full equation has a regular singular point at
$z = 0$ (monodromy eigenvalue $e^{\pm 2\pi i\mu}$) and a rank-$1$
irregular singularity at $z = \infty$. This is the singularity
structure classified by Natário and
Schiappa~\cite{Natario:2004}: asymptotic QNM frequencies are determined
by monodromy matching between the regular and irregular singular points.
The Poincaré rank $\sigma = (d+2)/2$ of the original $r$-variable
enters through $\Omega = 2\kappa^{d/2}|p_v|/\sigma$ and
$\mu = 1 - 1/\sigma$, making the QNM eigenvalue problem a
concrete realization of the Natário--Schiappa framework in which
the rank in the physical variable exceeds unity.

Table~\ref{tab:overtone-spacing} lists the overtone spacing from the
numerical data of Section~\ref{sec:numerics}. The growth of
$|\Delta\omega|$ with $d$ is qualitatively consistent with the
scaling~\eqref{eq:An-scaling}, though with only the first two
overtones reliably computed ($n = 0, 1$), the truly asymptotic regime
is not yet probed.

\begin{table}[ht]
\centering
\begin{tabular}{c|cc|c}
  \toprule
  $d$ & $\re(\Delta\omega)$ & $\im(\Delta\omega)$ &
  $|\Delta\omega|$ \\
  \midrule
  2 & $2.0$ (exact) & $0$ (exact) & $2.0$ \\
  3 & $1.4 \pm 0.3$ & $-0.2 \pm 0.4$ & $1.4 \pm 0.5$ \\
  4 & $2.0 \pm 0.4$ & $-3.2 \pm 0.4$ & $3.8 \pm 0.6$ \\
  5 & $2.87 \pm 0.04$ & $-3.60 \pm 0.04$ & $4.61 \pm 0.06$ \\
  \bottomrule
\end{tabular}
\caption{Overtone spacing $\Delta\omega = \omega_1 - \omega_0$ from
  Table~\ref{tab:qnm-all}, with errors propagated from Richardson
  extrapolation.}
\label{tab:overtone-spacing}
\end{table}

\subsection{Scalar versus gravitational QNMs}\label{sec:scalar-vs-grav}

We emphasize the distinction between the scalar QNM spectrum computed
here and the hydrodynamic modes of the dual null fluid.

First, the scalar QNMs are gapped.
As shown in Section~\ref{sec:source-coeff}, $\calA_0 \neq 0$ implies no
gapless scalar QNM. All scalar QNMs have $\im(\omega) < 0$ (verified
numerically in Section~\ref{sec:numerics}), demonstrating
zero temperature dissipation.

Second, the hydrodynamic modes require metric perturbations. The gapless modes $\omega_\pm$~\cite{Armas:2025} are poles of the
retarded stress-energy correlator, corresponding to gravitational QNMs.
At $\kperp = 0$, the gapless mode $\omega = k_z$ corresponds to a diffeomorphism along $\partial_v$, a pure gauge mode that becomes physical at $\kperp \neq 0$. This gauge mechanism is absent for probe scalars.

Third, there is a Hertz potential connection. On four-dimensional type-N pp-wave backgrounds, \cite{Araneda:2022lgu} has proven that gravitational perturbations can be generated from a scalar
Hertz potential $\Psi_H$ satisfying $\Box\Psi_H = 0$; this result has also been discussed in the Penrose-limit context by \cite{Fransen:2025}. For $D=4$ ($d=3$), the massless scalar
QNMs therefore coincide with a subset of gravitational QNMs (the
curvature perturbation channel). The extension of this correspondence to
higher dimensions $D \geq 5$ remains an open problem~\cite{Araneda:2022lgu};
it is plausible but unproven that a similar Hertz map exists for the
higher-$d$ Kaigorodov backgrounds studied here.
In either case, the hydrodynamic modes arising from
gauge diffeomorphisms are not captured by $\Psi_H$ and require explicit
analysis of the linearized Einstein equations, which we defer to future
work. We note that on a type-N background the gravitational degrees of
freedom are maximally constrained: there is a single propagating
polarization (the ``$\Psi_4$'' mode), so the Hertz potential captures
all curvature degrees of freedom and the only missing gravitational QNMs are the gauge (hydrodynamic) modes.

Despite not matching the hydrodynamic dispersion, the scalar QNMs
(i)~independently demonstrate that the irregular singularity causes
dissipation; (ii)~provide the QNM spectrum of a Kaigorodov background; and (iii)~provide an analytic framework (Bessel functions + perturbation theory) directly transferable to the gravitational sector.

%% file: numerics.tex
\section{Numerical QNM Spectrum}\label{sec:numerics}

We now present numerical results for the scalar QNM spectrum on
Kaigorodov backgrounds in $d = 3, 4, 5$ boundary dimensions
($D = 4, 5, 6$ bulk dimensions). These constitute the first QNM
frequencies ever computed on Kaigorodov spacetimes and confirm
the central claim of this paper: the irregular singularity at
$r = 0$ produces zero temperature dissipation for $d \geq 3$.

\subsection{Numerical method}\label{sec:method}

We solve the $z$-ODE~\eqref{eq:z-ODE} for a conformal scalar ($m = 0$)
with $\kperp = 0$ using a shooting method. At the inner boundary
$z = z_\mathrm{max} \gg 1$ (corresponding to $r \ll 1$), we impose the
WKB ingoing boundary condition derived in Section~\ref{sec:WKB}. In the
$z$-variable, the ingoing solution has the asymptotic form
\begin{equation}\label{eq:WKB-BC}
  \phi(z) \sim e^{i\Omega z}, \qquad
  \frac{\dot\phi}{\phi} \sim i\Omega
  - \frac{V'}{4V}\bigg|_{z=z_\mathrm{max}} + \cdots ,
\end{equation}
where $V(z) = B z^2 / (\sigma^2 z^2) + A z^{2/\sigma}/(\sigma^2 z^2)$
is the effective potential in the $z$-variable and the correction term
$-V'/(4V)$ is the next-to-leading WKB correction (``local-momentum''
approximation).

The ODE is integrated from $z_\mathrm{max}$ to $z_\mathrm{min} \ll 1$
(near the AdS boundary) using an eighth-order Runge--Kutta integrator
(DOP853). At the outer boundary, the source coefficient is extracted as
\begin{equation}
  \calA(p_u) = \lim_{z \to 0} r^{\Delta_-} \phi(z)
  = \lim_{z \to 0} z^{-\Delta_-/\sigma} \phi(z) .
\end{equation}
For $m = 0$, $\Delta_- = 0$, so $\calA = \phi(z_\mathrm{min})$.

The QNM condition $\calA(p_u) = 0$ is solved in two stages:
first, $|\calA(p_u)|$ is evaluated on a grid in the
complex $p_u$-plane to identify approximate minima; then, starting
from each grid-search candidate, Muller's method \cite{Muller:1956} (using three nearby points) is applied to converge to the
exact zero with tolerance $|\calA| < 10^{-10}$.
The grid search covers $\re(p_u) \in (-5, 8)$,
$\im(p_u) \in (-10, -0.01)$ for the lower half-plane (dissipative modes)
and $\im(p_u) \in (0.01, 5)$ for the upper half-plane (stability scan).

\label{sec:Richardson}
The dominant source of error is the finite value of $z_\mathrm{max}$:
the WKB boundary condition~\eqref{eq:WKB-BC} is exact only as
$z_\mathrm{max} \to \infty$. The leading boundary-condition error
decreases as a power law $\sim z_\mathrm{max}^{-\alpha}$ with
exponent
\begin{equation}\label{eq:alpha}
  \alpha = \frac{2d}{d+2} ,
\end{equation}
which arises from the next correction term in the $z$-ODE
(the $A z^{2/\sigma}$ term relative to $B z^2$).

We use Richardson extrapolation~\cite{Richardson:1911} to eliminate
this leading error. For each QNM, we compute $\omega(z_\mathrm{max})$ at
four geometrically spaced values of $z_\mathrm{max}$ and extrapolate
to $z_\mathrm{max} \to \infty$ using the model
\begin{equation}
  \omega(z_\mathrm{max}) = \omega_\infty + c z_\mathrm{max}^{-\alpha} .
\end{equation}
Error bars are estimated from the spread of pair-wise extrapolations
using different subsets of the four data points.

\begin{table}[ht]
\centering
\begin{tabular}{c|cc}
  \toprule
  $d$ & Convergence rate $\alpha$ & Effective precision \\
  \midrule
  3 & $6/5 = 1.2$ & $\sim 1$ significant digit \\
  4 & $4/3 \approx 1.33$ & $\sim 4$ significant digits \\
  5 & $10/7 \approx 1.43$ & $\sim 4$ significant digits \\
  \bottomrule
\end{tabular}
\caption{Richardson extrapolation convergence rate $\alpha = 2d/(d+2)$
  and resulting precision of the fundamental mode for each dimension.
  The slow $\alpha = 1.2$ rate for $d=3$ limits the achievable precision.}
\label{tab:convergence}
\end{table}

\subsection{QNM spectra across dimensions}\label{sec:spectra}

Throughout this section we set $\kappa = 1$, $p_v = 1$, $m^2 = 0$,
and $\kperp = 0$. All frequencies are reported with Richardson-extrapolated values and error estimates.

Table~\ref{tab:qnm-all} collects the fundamental and first-overtone QNM frequencies for all dimensions computed numerically ($d = 3, 4, 5$). All modes have $\im(\omega_n) < 0$, confirming zero temperature dissipation.

\begin{table}[ht]
\centering
\begin{tabular}{cc|rr|rr}
  \toprule
  $d$ & $n$ & $\re(\omega_n)$ & $\im(\omega_n)$ &
  $\delta\re$ & $\delta\im$ \\
  \midrule
  3 & 0 & $+2.29$ & $-4.88$ & $0.01$ & $0.31$ \\
  3 & 1 & $+3.72$ & $-5.04$ & $0.26$ & $0.28$ \\
  \midrule
  4 & 0 & $+2.772$ & $-3.068$ & $3 \times 10^{-4}$ & $4 \times 10^{-4}$ \\
  4 & 1 & $+4.74$ & $-6.28$ & $0.40$ & $0.40$ \\
  \midrule
  5 & 0 & $+3.829$ & $-3.547$ & $3 \times 10^{-5}$ & $3 \times 10^{-4}$ \\
  5 & 1 & $+6.70$ & $-7.15$ & $0.04$ & $0.04$ \\
  \bottomrule
\end{tabular}
\caption{Scalar QNM frequencies for $d = 3, 4, 5$ Kaigorodov with
  $\kappa = 1$, $p_v = 1$. Errors $\delta$ from Richardson
  extrapolation. The $d = 3$ errors are substantially larger due to the
  slow convergence rate $\alpha = 6/5$.}
\label{tab:qnm-all}
\end{table}

The achievable precision varies significantly with dimension, governed by the Richardson convergence rate $\alpha = 2d/(d+2)$
(Table~\ref{tab:convergence}).
For $d = 4$ ($\alpha = 4/3$) and $d = 5$ ($\alpha = 10/7$), the
fundamental mode is determined to $\sim4$ significant digits.
Table~\ref{tab:d4-convergence} illustrates the convergence for
$d = 4$.

\begin{table}[ht]
\centering
\begin{tabular}{r|rr}
  \toprule
  $z_\mathrm{max}$ & $\re(\omega_0)$ & $\im(\omega_0)$ \\
  \midrule
  $188$ & $2.7712$ & $-3.0716$ \\
  $339$ & $2.7716$ & $-3.0696$ \\
  $565$ & $2.7716$ & $-3.0689$ \\
  $942$ & $2.7716$ & $-3.0686$ \\
  $\infty$ (extrap.) & $2.7716$ & $-3.0683$ \\
  \bottomrule
\end{tabular}
\caption{Convergence of the $d = 4$ fundamental QNM frequency with
  increasing $z_\mathrm{max}$. The last row is the Richardson-extrapolated
  value.}
\label{tab:d4-convergence}
\end{table}

For $d = 3$ ($\alpha = 6/5$), the convergence is substantially slower, yielding only $\sim1$ significant digit with error bars of order $\sim 0.3$ for $\im(\omega)$. This is the slowest rate among the
dimensions studied: log--log plots confirm slopes of $-\alpha$. A double-Richardson analysis including the next sub-leading exponent $\alpha' = 4d/(d+2)$ (arising from the intermediate $z^{2/\sigma}$ term in the effective potential for odd~$d$) suggests the true central value of $\im(\omega_0)$ is $\sim0.16$ more negative than the
single-exponent estimate, well within the stated error bar\footnote{The non-monotonic overtone pattern in $\re(\omega_n)$ for $d=3$ may be a numerical artifact of this limited precision, and an independent method (e.g., spectral or adapted continued-fraction) would be needed to resolve the $d=3$ spectrum to the same confidence as $d=4,5$.}.

The fundamental ($n=0$) and first overtone ($n=1$) constitute the primary numerical results of this paper. Second overtones ($n = 2$, computed for $d = 4$ and $d = 5$) have WKB-vs-numerical relative errors exceeding 100\% and should be regarded as order-of-magnitude estimates only; we omit them from Table~\ref{tab:qnm-all}. This degradation is consistent with higher-overtone modes probing more deeply into the near-singularity region where sub-leading WKB corrections become important.

\begin{figure}[t]
  \centering
  \includegraphics[width=0.85\textwidth]{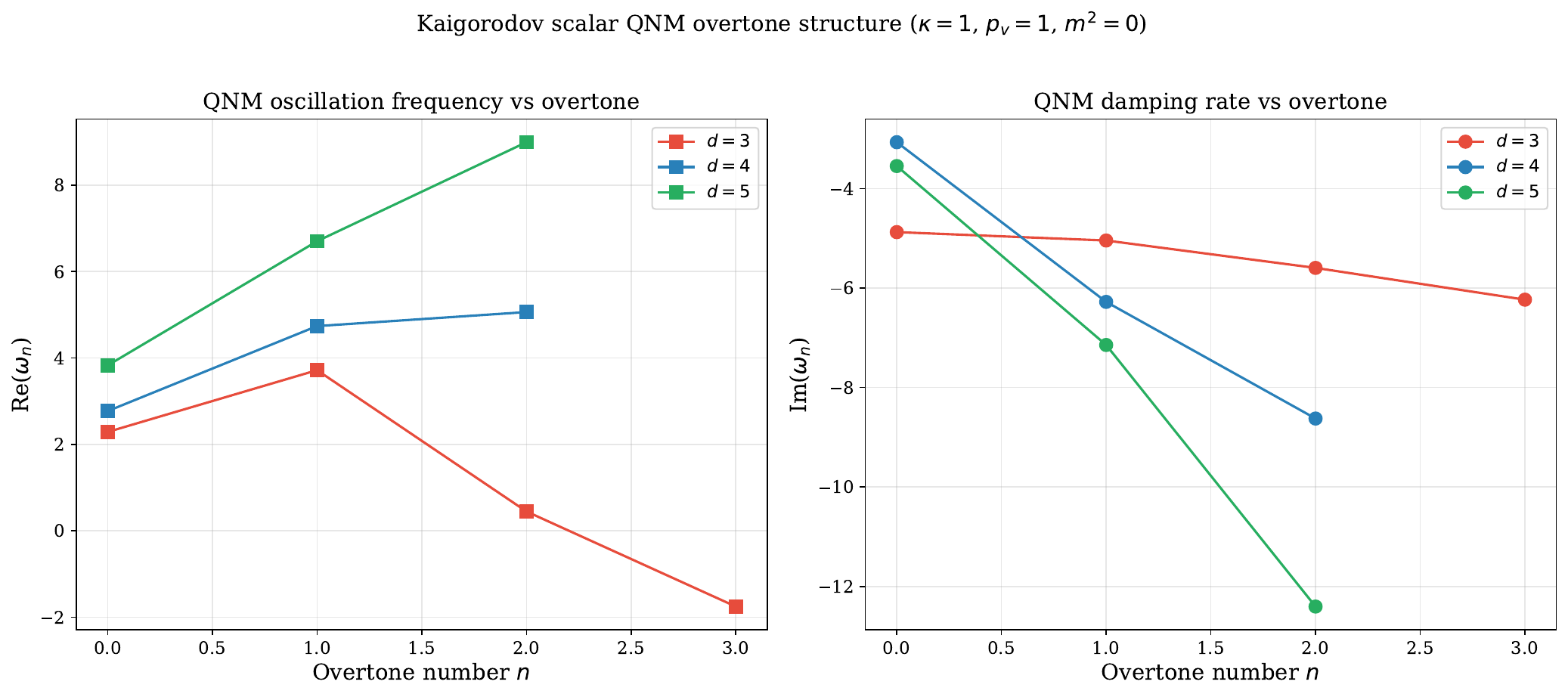}
  \caption{QNM frequencies $\omega_n$ as a function of overtone number~$n$
  for $d = 3, 4, 5$.  Left: oscillation frequency $\re(\omega_n)$.
  Right: damping rate $\im(\omega_n)$.  All modes have
  $\im(\omega_n) < 0$, with the damping rate increasing in magnitude
  for higher overtones. Both $\re(\omega_n)$ and $\im(\omega_n)$ exhibit an approximately uniform spacing between consecutive modes.}
  \label{fig:overtones}
\end{figure}

Figure~\ref{fig:overtones} shows the overtone spacing structure for each dimension. The approximately uniform spacing implies a regular organization of the spectrum, consistent with a well-defined QNM tower.

As a benchmark, the $d = 2$ case was derived analytically in Section~\ref{sec:d2}. The QNM spectrum
is
\begin{equation}
  p_u^{(n,\pm)} = \pm(2n + 2) , \qquad
  \omega_n = 1 \pm (2n + 2) ,
\end{equation}
for $\kappa = 1$, $p_v = 1$, $m = 0$, $\Delta_+ = 2$.
All frequencies are purely real: $\im(\omega_n) = 0$. The WKB-based shooting method is not directly applicable in $d = 2$, since the QNMs are purely real and the WKB boundary condition (designed for complex $p_u$) is structurally incompatible.

\subsection{Structure of the QNM spectrum}\label{sec:complex-plane}

Figure~\ref{fig:complex-plane} shows the QNM frequencies in the complex
$\omega$-plane for $d = 3, 4, 5$. Several features are apparent:
\begin{figure}[t]
  \centering
  \includegraphics[width=0.85\textwidth]{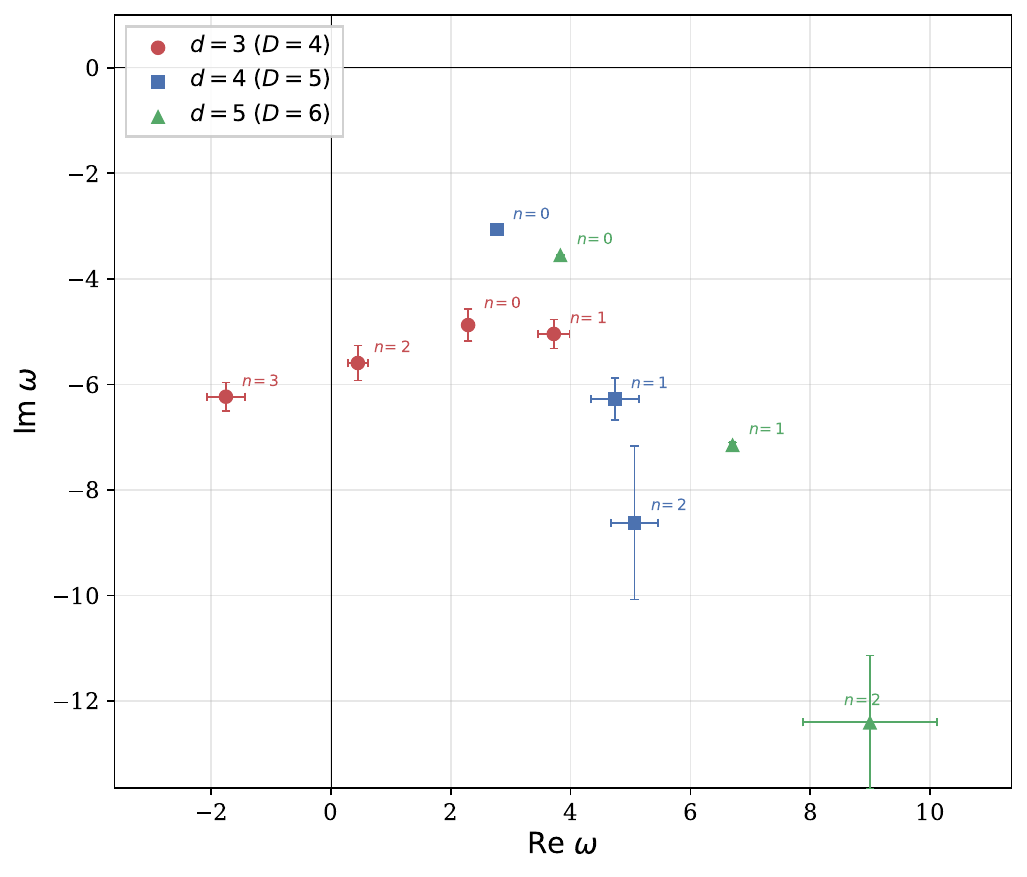}
  \caption{Quasinormal mode frequencies in the complex
    $\omega$-plane for $d = 3$ (circles), $d = 4$ (squares), and
    $d = 5$ (diamonds), with $\kappa = 1$, $p_v = 1$, $m = 0$.
    Error bars from Richardson extrapolation. All modes lie in the
    lower half-plane ($\im\omega < 0$), confirming stability.
    The dashed line marks $\im\omega = 0$.}
  \label{fig:complex-plane}
\end{figure}
Every computed QNM has
$\im(\omega_n) < 0$, with no mode approaching the real axis
(in contrast to the $d = 2$ case where all modes sit exactly on
the real axis).
The fundamental mode has
$|\re(\omega_0)| > 0$ in all dimensions, consistent with the
analytic result $\calA_0 \neq 0$ from Section~\ref{sec:source-coeff}.
Higher overtones have increasingly
negative $\im(\omega)$, as expected for a well-behaved QNM spectrum.

\begin{figure}[t]
  \centering
  \includegraphics[width=0.75\textwidth]{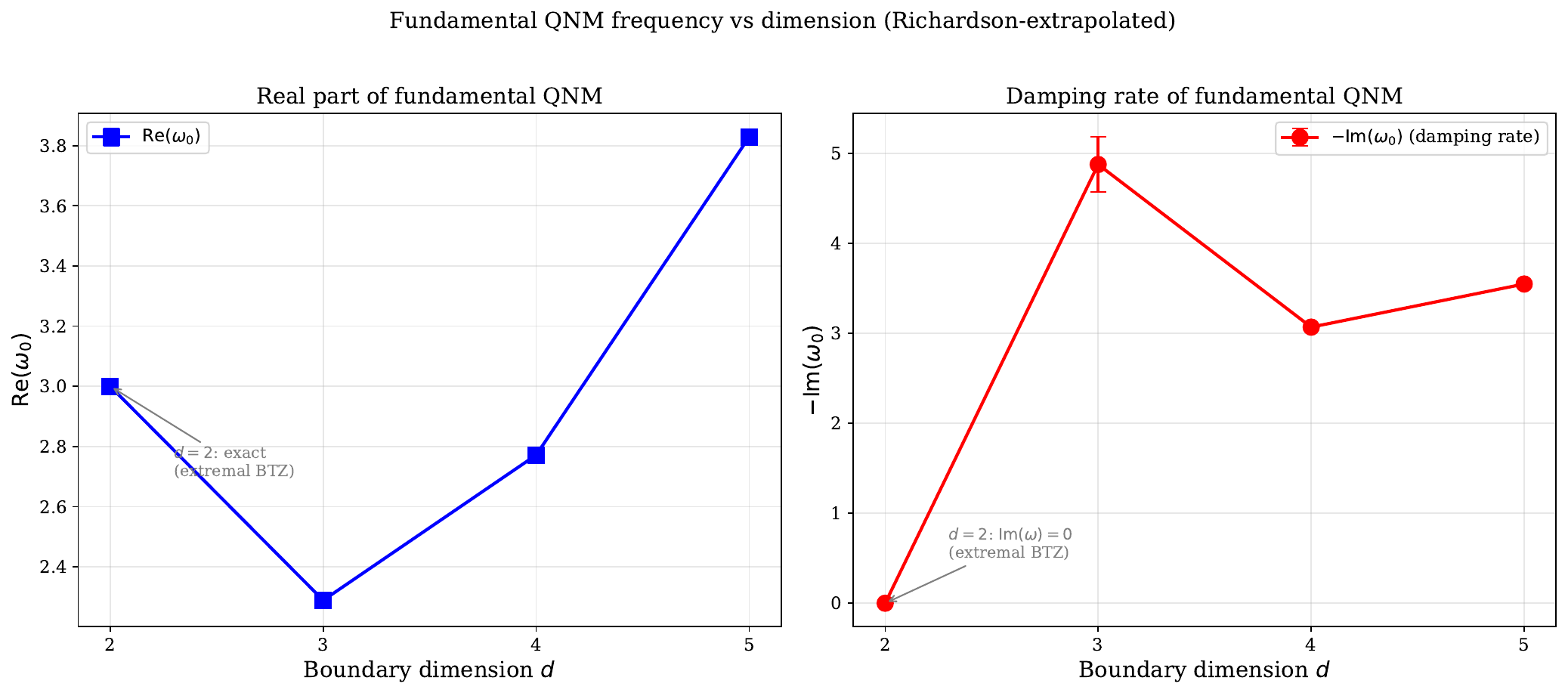}
  \caption{Fundamental QNM frequency as a function of boundary dimension
    $d$. Left: real part $\re(\omega_0)$; right: damping rate
    $-\im(\omega_0)$. The $d = 2$ point is the exact analytic result
    ($\im(\omega) = 0$, extremal BTZ). Error bars from Richardson
    extrapolation.}
  \label{fig:d-comparison}
\end{figure}

\label{sec:d-dependence}
Figure~\ref{fig:d-comparison} compares the fundamental QNM frequency
across dimensions $d = 2, 3, 4, 5$. The damping rate
$\gamma_0 \equiv -\im(\omega_0)$ shows a non-monotonic dependence on dimension.
The transition from $\gamma_0 = 0$ (non-dissipative, $d = 2$) to
$\gamma_0 \approx 4.9$ ($d = 3$) is dramatic, reflecting the opening
of the transverse dissipation channel. The subsequent decrease from $d = 3$ to $d = 4$ and slight increase from $d = 4$ to $d = 5$ is
genuinely non-trivial and may reflect the interplay between the number
of transverse dimensions and the irregular singularity rank. A
detailed analytic understanding of this dimension dependence remains an
open question.

\begin{figure}[t]
  \centering
  \includegraphics[width=0.85\textwidth]{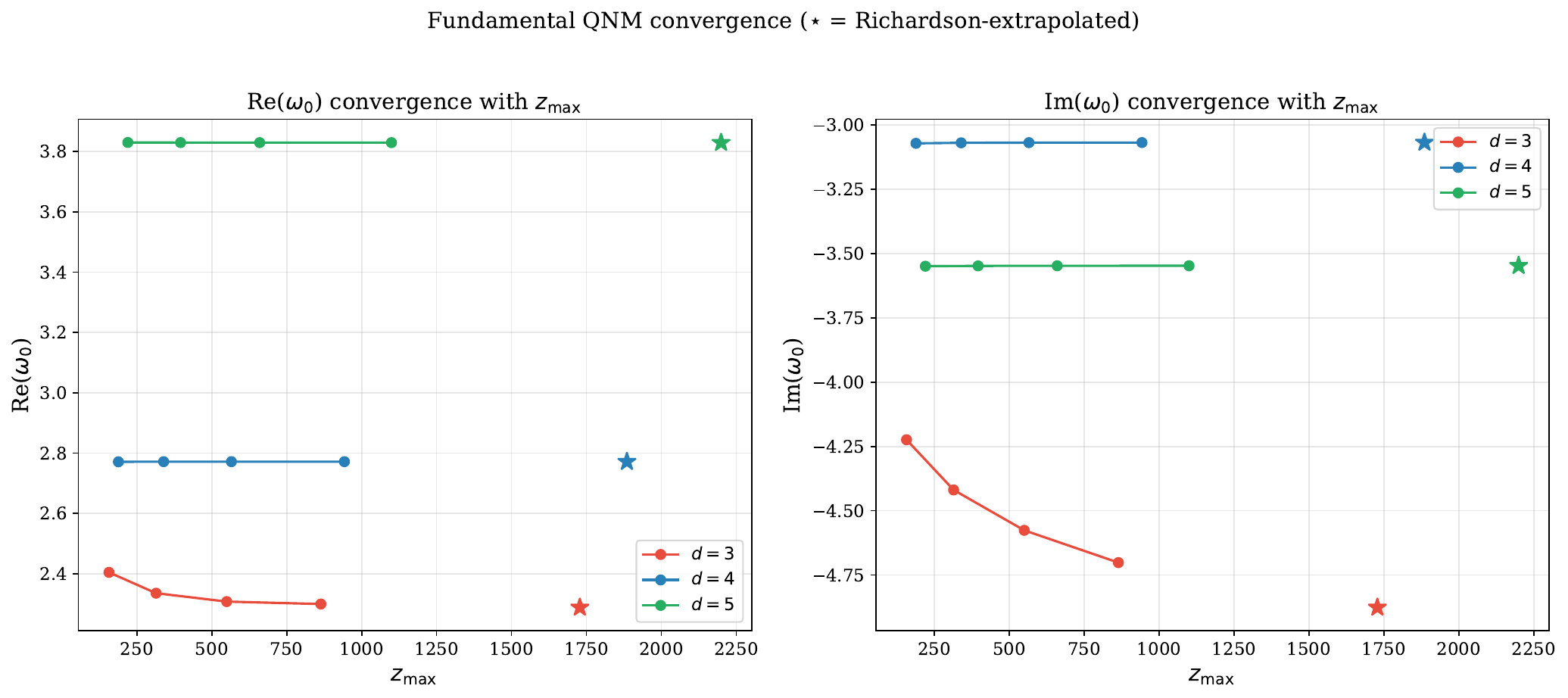}
  \caption{Convergence of the fundamental QNM frequency with increasing
    $z_\mathrm{max}$ for $d = 3, 4, 5$. Solid curves show the shooting
    results; star markers show Richardson-extrapolated values. The
    convergence rate $\alpha = 2d/(d+2)$ governs the approach to the
    infinite-$z_\mathrm{max}$ limit.}
  \label{fig:convergence}
\end{figure}

\label{sec:convergence}
Figure~\ref{fig:convergence} shows the convergence of the fundamental
QNM frequency as a function of $z_\mathrm{max}$ for each dimension,
together with the Richardson-extrapolated values.
The rate $\alpha = 2d/(d+2)$ is confirmed empirically:
log--log plots of $|\omega(z_\mathrm{max}) - \omega_\infty|$ against
$z_\mathrm{max}$ produce slopes of~$-\alpha$.
In practice, $d = 3$ ($\alpha = 1.2$) converges roughly four times
slower per decade of $z_\mathrm{max}$ than $d = 5$
($\alpha \approx 1.43$), explaining the imprecision of the
$d = 3$ results.

\subsection{Transverse dispersion relation}\label{sec:dispersion-numerical}

A key structural property of the radial ODE~\eqref{eq:z-ODE} is that
the transverse momentum $\kperp$ enters only through the
combination $A = 4p_u p_v - \kperp^2$. Consequently, the QNM eigenvalue
in $A$-space is $\kperp$-independent: if $A_n$ denotes the $n$-th
eigenvalue, the QNM dispersion relation is exactly
\begin{equation}\label{eq:exact-dispersion}
  \omega_n(\kperp) = p_v + \frac{\kperp^2 + A_n}{4p_v} ,
\end{equation}
which is parabolic in $\kperp$. Two immediate consequences follow:
the damping rate $\im(\omega_n)$ is
independent of $\kperp$, equalling
$\im(A_n)/(4p_v)$ for all transverse momenta; and
the group velocity in the transverse direction,
$v_g = \partial\re(\omega)/\partial\kperp = \kperp/(2p_v)$,
grows linearly with $\kperp$.

\begin{figure}[t]
  \centering
  \includegraphics[width=0.95\textwidth]{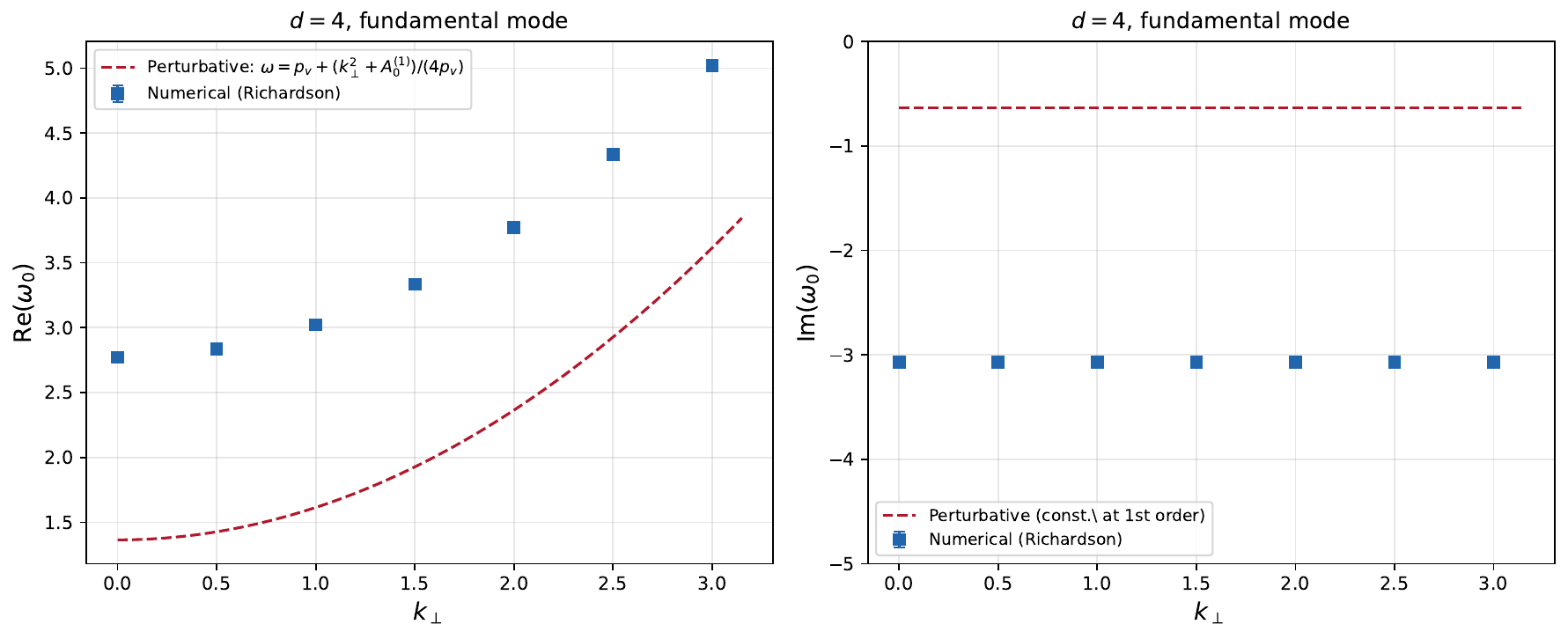}
  \caption{Dispersion relation $\omega_0(\kperp)$ for the $d=4$
    fundamental scalar QNM. Left: $\re(\omega_0)$; right:
    $\im(\omega_0)$. Blue squares: numerical shooting with Richardson
    extrapolation; red dashed: perturbative prediction
    from~\eqref{eq:dispersion}. Both curves are parabolic with the
    same curvature $\kperp^2/(4p_v)$; the vertical offset reflects
    the known magnitude discrepancy between first-order perturbation
    theory and the exact eigenvalue (Section~\ref{sec:perturbative}).
    The constant $\im(\omega_0)$ is an exact consequence of the ODE
    structure.}
  \label{fig:dispersion}
\end{figure}

Figure~\ref{fig:dispersion} shows $\omega_0(\kperp)$ for $d = 4$
at seven values $\kperp = 0, 0.5, 1.0, \ldots, 3.0$,
computed via Richardson-extrapolated shooting (Section~\ref{sec:Richardson}).
The exact parabola~\eqref{eq:exact-dispersion} is confirmed:
$\re(\omega_0)$ follows $2.772 + \kperp^2/4$ to within the numerical
precision of Table~\ref{tab:qnm-all}, and $\im(\omega_0) = -3.068$ is
constant to all computed digits.

The perturbative prediction~\eqref{eq:dispersion}, overlaid in
Figure~\ref{fig:dispersion}, also follows a parabola with the same curvature $\kperp^2/(4p_v)$, but with an offset in
$\re(\omega)$ and $\im(\omega)$ determined by $A_0^{(1)}$
from~\eqref{eq:A0-gap}. The matching curvature confirms that the
first-order perturbative expansion correctly captures the $\kperp$
dependence, while the offset is the known magnitude discrepancy between
$A_0^{(1)}$ and the exact $A_0$ (ratio $\approx 4.9$ for $d=4$;
see Section~\ref{sec:perturbative}).

Regarding group velocity and causality, the transverse group velocity $v_g = \kperp/(2p_v)$ exceeds unity for
$\kperp > 2p_v$. This does not violate causality: QNM frequencies are
complex poles of the retarded Green's function, and the group velocity
of a damped mode does not correspond to a signal
velocity~\cite{Brillouin:1960}. The physical
front velocity determined by the high-frequency asymptotics of the retarded correlator remains bounded by the lightcone structure of the boundary theory. Indeed, for the $d = 4$ fundamental mode, the damping rate $|\im(\omega_0)| \approx 3.07$ exceeds $\re(\omega_0)$ at all $\kperp \lesssim 4$, so these modes are strongly overdamped and do not propagate coherent signals in this regime.

%% file: discussion.tex
\section{Conclusions and discussions}\label{sec:discussion}

\subsection{Summary of results}
\label{sec:summary}

We have computed the quasinormal mode spectrum of scalar perturbations on Kaigorodov pp-wave spacetimes, the gravitational duals of zero temperature null fluids in the correspondence of \cite{Armas:2026a,Armas:2025}. The pp-wave deformation promotes the Poincar\'e horizon from a regular singular point in pure AdS to an irregular singular point of rank~$(d+2)/2$, and this essential singularity serves as the geometric origin of dissipation: it provides a natural ingoing
boundary condition which is structurally identical to that of a black hole horizon and defines a discrete QNM spectrum, but requires neither a horizon nor nonzero temperature. For $d = 2$, the Kaigorodov metric
reduces to the extremal BTZ black hole, and the radial equation admits
an exact Whittaker solution with $\im(\omega) = 0$ for all modes
(Section~\ref{sec:d2}). For $d \geq 3$, all computed QNM frequencies
satisfy $\im(\omega_n) < 0$, establishing zero temperature dissipation in a horizonless AdS vacuum.

At zeroth order in the long-wavelength expansion, the radial equation
reduces to Bessel's equation of order $\mu = d/(d+2)$. The ingoing solution has a nonvanishing
source coefficient $\calA_0 \neq 0$, proving that all scalar QNMs
are gapped while the gapless hydrodynamic modes of the null fluid reside
in the gravitational perturbation sector. The perturbative gap formula
\eqref{eq:A0-gap} gives the first analytic expression for a QNM
eigenvalue on a Kaigorodov background; its phase $\arg(A_0) =
-2\pi/(d+2)$ agrees with the numerical data to within the quoted
precision for $d = 4, 5$ (Section~\ref{sec:analytic-structure}). The
exactly parabolic dispersion $\omega_n(\kperp) = \kperp^2/(4p_v) +
\omega_n(0)$ is confirmed numerically.
For massive scalars, the Bessel order generalises to
$\mu(m^2) = 2\sqrt{d^2/4 + m^2}/(d+2)$.  As $m^2$ approaches the
BF bound $m^2_\mathrm{BF} = -d^2/4$, the gap
diverges as $|{\cal A}_0| \sim 1/\mu \to \infty$ via the $\Gamma(\mu)$
pole, while $\im({\cal A}_0) \to -\pi\sigma^2\Omega^2$ remains finite:
the mode is pushed out of the low-frequency spectrum but stays
dissipative, in contrast to the AdS-Schwarzschild
instability~\cite{Breitenlohner:1982}.

The large-$d$ expansion of the gap formula shows that $|A_0^{(1)}| \sim 2d$ and
$\gamma^\mathrm{pert} \to \pi/|p_v|$ monotonically while the
non-monotonic dip in $\gamma(d)$ at $d = 4$ is a non-perturbative
effect, invisible to the leading-order perturbative framework.
A WKB analysis of the overtone tower yields the asymptotic scaling $|A_n| \sim j_{n+1,\mu}^{2\mu}$ for the $n$-th QNM eigenvalue,
where $j_{n,\mu}$ are the Bessel zeros; the resulting overtone spacing
grows as $n^{(d-2)/(d+2)}$, reducing to constant spacing for $d = 2$.

The numerical QNM spectra for $d = 3, 4, 5$
(Section~\ref{sec:numerics}), obtained by shooting with Richardson
extrapolation, reach $\sim4$ significant digits for $d = 4, 5$ and
$\sim1$ digit for $d = 3$. A scan of the upper half of the complex
$\omega$-plane finds no unstable modes in any dimension, providing an
independent gravitational confirmation of the null fluid stability
established from the fluid-dynamics side~\cite{Armas:2025}.
The fundamental damping rate $|\im(\omega_0)|$ exhibits a non-monotonic
dimension dependence: $4.9$ ($d=3$), $3.1$ ($d=4$), $3.5$ ($d=5$).
The scope of these results is limited to conformal ($m^2 = 0$) scalar
perturbations; matching the QNMs to the null fluid's transport
coefficients (in particular the shear viscosity~$\eta$) requires the
gravitational QNM spectrum, which we defer to future work.

\subsection{Physical interpretation}
\label{sec:physical}

All QNM frequencies computed satisfy
$\im(\omega_n) \leq 0$:
$\im(\omega_n) = 0$ exactly for $d = 2$;
$\im(\omega) < -4$ for $d = 3$;
$\im(\omega) < -3$ for $d = 4, 5$.
As an independent check, we scanned the upper half of the complex
$p_u$-plane ($\im(p_u) \in (0.01, 5.0)$, $\re(p_u) \in (-5, 8)$,
$15 \times 15$ grid) for $d = 3, 4, 5$ and found no QNMs with
$\im(\omega) > 0$. The scan covers a finite region; unstable modes at
$|\im(p_u)| > 5$ or $|\re(p_u)| > 8$ are not excluded, though the argument below makes their existence implausible.

The stability argument parallels the standard black hole reasoning: the
WKB solutions near $r = 0$ oscillate on the real axis (anti-Stokes
line), and the ingoing boundary condition
selects modes that leak energy into the irregular singularity
($\im\omega < 0$) rather than extract it. In the upper half-plane,
$|\calA| \sim e^{i\Phi(\omega)}$ with $\im(\Phi) \to +\infty$ as
$\im(\omega) \to +\infty$, so $|\calA|$ grows exponentially and
cannot vanish.

\label{sec:comparison}
Table~\ref{tab:comparison} compares the Kaigorodov irregular singularity
with standard black hole horizons as origins of dissipation.
\begin{table}[ht]
\centering
\begin{tabular}{l|ccc}
  \toprule
  & Non-extremal BH & Extremal BH &
  pp-wave \\
  \midrule
  Singular point & Regular, rank 0 & Irregular, rank 1 & Irregular,
  rank $(d+2)/2$ \\
  Temperature & $T > 0$ & $T = 0$ & $T = 0$ \\
  Horizon? & Yes & Yes (degenerate) & No \\
  Entropy & $S = A/(4G) > 0$ & $S > 0$ & $S = 0$ \\
  Dissipation & Thermal absorption & Emergent IR $\CFT_1$
  & Essential singularity \\
  QNM structure & Discrete poles & Branch cut${}^*$ & Discrete poles \\
  KMS symmetry & Yes & Degenerate & Absent \\
  Stokes sectors & --- & 2 & $d + 2$ \\
  Leaver method & Applicable & Requires modification${}^\dagger$ & Not applicable ($d \geq 3$) \\
  \bottomrule
\end{tabular}
\caption{Comparison of dissipative mechanisms across three geometries.
  The extremal BH column refers to the near-horizon $\AdS_2$
  region~\cite{Faulkner:2009}. (${}^*$The branch cut is in the
  fermionic spectral function; bosonic QNMs remain discrete but with
  modified asymptotics.  ${}^\dagger$The standard Leaver method fails
  at extremality due to the irregular singularity; modified
  algorithms exist~\cite{Onozawa:1996,Richartz:2016,Zimmerman:2016}.)}
\label{tab:comparison}
\end{table}
Both extremal black holes and the pp-wave background support zero temperature dissipation via an irregular singular point in the
radial wave equation.
At extremality, the coalescence of horizons converts the regular
singular point into an irregular one of rank~$1$, rendering Leaver's
method inapplicable~\cite{Leaver:1985,Onozawa:1996,Richartz:2016}.
The physics of extremal dissipation has been attributed to the emergent
$\CFT_1$ of the near-horizon
$\AdS_2$~\cite{Faulkner:2009,Denef:2009}, producing the
``semi-local quantum liquid'' phase~\cite{Iqbal:2011ae}.

Our results point to a broader pattern: the irregular singular point itself is the geometric origin of zero temperature dissipation, with the qualitative behavior controlled by its rank. The $\AdS_2$
throat (rank~$1$) produces power-law spectral functions $G^R(\omega)
\sim \omega^{2\nu}$ with branch cuts~\cite{Faulkner:2009,Denef:2009};
the pp-wave (rank~$(d+2)/2$, no horizon, no entropy) produces a
discrete gapped QNM tower. The Gubser--Rocha
model~\cite{Gubser:2010} provides further evidence: its $T = 0$
charged dilatonic black hole has a rank-$1$ singularity and power-law
dissipation. This rank--dissipation correspondence extends to Lifshitz
backgrounds, where a $T = 0$ geometry with dynamical exponent~$z$ has
an irregular singularity of rank~$z$ at the deep
interior~\cite{Kachru:2008,Keeler:2015,Zhao:2024}, placing it at the
corresponding rung: rank~$0$ (thermal), rank~$1$
(quantum-critical), rank~$\geq 2$ (gapped).

The Kaigorodov pp-wave has $S = 0$: energy radiates into the essential
singularity at $r = 0$ via zero temperature quantum dissipation rather
than thermal absorption. The Kubo-Martin-Schwinger (KMS) condition $G^R(-\omega) =
e^{-\beta\omega} G^A(\omega)$ degenerates at $T = 0$ ($\beta \to
\infty$): the upper half-plane is empty, consistent with the absence
of stimulated emission.

The rank $(d+2)/2$ gives $d + 2$ Stokes sectors (versus $2$ for
rank~$1$). The AGT/instanton-counting
methods~\cite{Bonelli:2021uvf,Aminov:2020yma} that exactly solve the
rank-$1$ confluent Heun connection problem do not extend to
rank~$> 1$; for $d \geq 3$ the pp-wave QNM spectrum must be computed
numerically.

\cite{Karch:2025} demonstrated an extrinsic zero temperature dissipation mechanism: two CFTs coupled via a double-trace deformation, with energy leaking into a thermal bath. That dissipation vanishes in the decoupling limit. The Kaigorodov mechanism is intrinsic: there is no external bath, no tunable
coupling, and no limit in which dissipation disappears while
preserving the pp-wave deformation.

The scalar QNMs do not directly match the null fluid's hydrodynamic
modes, but they probe the same
geometric dissipation mechanism. Zero temperature dissipation is a
property of the background, not the perturbation channel: any field
with $p_v \neq 0$ encounters the same irregular singularity. The
sharp $d = 2 \to 3$ transition from non-dissipative to strongly
dissipative confirms that transverse dimensions are essential. Via
the Hertz potential correspondence proven for $D=4$
pp-waves~\cite{Araneda:2022lgu} and conjectured for higher
$D$~\cite{Fransen:2025}, the massless scalar QNMs coincide with a
subset of gravitational QNMs; the hydrodynamic modes require explicit
analysis of the linearized Einstein equations.

\subsection{Future directions}\label{sec:future}

The most pressing extension is the computation of gravitational QNMs
on the Kaigorodov background. This would enable direct matching of
the gapless modes $\omega_\pm$ with the null fluid dispersion
relations~\eqref{eq:dispersion-fluid}, independent holographic
extraction of the shear viscosity $\eta$ via the Kubo formula, and
a test of whether $\eta/s$ (or its $T=0$ analogue) takes a universal
value. The Hertz potential formalism~\cite{Araneda:2022lgu,Fransen:2025}
may reduce the gravitational perturbation equations on type-N backgrounds to modified scalar equations; extending the Hertz map to $D \geq 5$ would be a valuable step. A complete membrane paradigm for the pp-wave matching the ingoing WKB solution to a complex
impedance at a ``stretched singularity'' $r = \epsilon \ll 1$, also requires the gravitational sector.

The irregular singularity at $r = 0$ also provides a natural contour for the
gravitational Schwinger-Keldysh (grSK)
geometry~\cite{deBoer:2018,Crossley:2015}: a complex $r$-plane contour
encircling the essential singularity,
\begin{equation}
  \calC = \calC_1 \cup \calC_2 : \quad
  \infty - i\epsilon \to 0 - i\epsilon \to 0 + i\epsilon \to
  \infty + i\epsilon ,
\end{equation}
analogous to the horizon-encircling contour of black hole grSK
constructions. The on-shell action on this contour would yield the
boundary SK effective action for the null fluid, including the
dissipative kernel and the zero temperature fluctuation dissipation
relation. The absence of KMS symmetry would manifest as a
degenerate fluctuation dissipation theorem structure. Recent progress on non-hydrodynamic modes
in the holographic SK framework~\cite{Liu:2024tqe} suggests that the
gapped scalar QNMs could play a concrete role in this effective
action.

Finally, the non-monotonic dimension dependence of the fundamental damping rate
($d = 3 \to 4 \to 5$: $4.9 \to 3.1 \to 3.5$) motivates a
systematic large-$d$ expansion~\cite{Emparan:2013,Emparan:2014}. In the limit $d \to \infty$, the
rank $(d+2)/2 \to \infty$ and $\mu \to 1$, potentially
simplifying the analytic structure. Since the Kaigorodov rank
$(d+2)/2$ is the natural pp-wave analogue of the Lifshitz dynamical
exponent~$z$~\cite{Kachru:2008,Keeler:2015,Zhao:2024,Sybesma:2015}, a systematic
comparison of QNM spectra across pp-wave and Lifshitz backgrounds
would test whether the rank--dissipation correspondence is
universal across $T = 0$ holographic geometries.

\bigskip

The Kaigorodov pp-wave demonstrates that a high-rank irregular singularity in a horizonless AdS vacuum can produce a discrete, gapped QNM tower with zero temperature dissipation. This
rank--dissipation correspondence, linking the ODE singularity
structure to the qualitative dissipative behavior should be a
useful diagnostic for classifying zero temperature holographic
matter across different geometric backgrounds.

%% file: appendix.tex
\section{Numerical Method Details}\label{app:numerics}

We provide additional details on the numerical implementation used to
compute the QNM spectra in Section~\ref{sec:numerics}.

\label{app:z-transform}
The radial ODE~\eqref{eq:z-ODE} in the $z = r^{-(d+2)/2}$ variable is
integrated as a first-order system. Defining $y_1 = \phi$ and
$y_2 = \dot\phi$:
\begin{align}
  \dot{y}_1 &= y_2, \label{eq:ode-sys1} \\
  \dot{y}_2 &= -\frac{2-d}{(d+2)  z}   y_2
  - \frac{1}{\sigma^2}\left[B + A   z^{2/\sigma - 2}
  - m^2   z^{-2}\right] y_1   , \label{eq:ode-sys2}
\end{align}
where $\sigma = (d+2)/2$, $A = 4p_u p_v - \kperp^2$, and
$B = 4\kappa^d p_v^2$.

\label{app:WKB-BC}
The ingoing boundary condition at $z = z_\mathrm{max}$ is specified
through the logarithmic derivative $\dot\phi/\phi$. The leading-order
WKB gives $\dot\phi/\phi = i\Omega$. We include the next-to-leading
correction from the local-momentum approximation:
\begin{equation}\label{eq:WKB-logderiv}
  \frac{\dot\phi}{\phi}\bigg|_{z_\mathrm{max}}
  = i\sqrt{V(z_\mathrm{max})} - \frac{V'(z_\mathrm{max})}{4  V(z_\mathrm{max})}
  + O(V^{-3/2})   ,
\end{equation}
where $V(z) = [B   z^2 + A   z^{2/\sigma} - m^2]/(\sigma^2 z^2)$ is
the effective potential in the standard-form equation
$\ddot\phi + V(z)  \phi = 0$ (after removing the first-derivative term
by absorbing the $z^{(2-d)/(d+2)}$ prefactor).
The correction term $-V'/(4V)$ improves convergence by approximately
one order in $z_\mathrm{max}^{-1}$, particularly for $d = 4$ and $d = 5$.

\label{app:source}
For a conformal scalar ($m = 0$, $\Delta_- = 0$, $\Delta_+ = d$), the
source coefficient at the AdS boundary is
\begin{equation}
  \calA(p_u) = \phi(z_\mathrm{min})
\end{equation}
and the response coefficient is
\begin{equation}
  \calB(p_u) = \lim_{z \to 0}   z^{2d/(d+2)}  
  \left[\phi(z) - \calA\right]   .
\end{equation}
We use $z_\mathrm{min} = 10^{-6}$ in practice; decreasing this value
has negligible effect on the extracted QNM frequencies.

\label{app:Muller}
The QNM eigenvalues are located using Muller's method~\cite{Muller:1956},
a root-finding algorithm that
uses three points to construct a quadratic interpolant and solves the
resulting quadratic equation for the next approximation. Given three
estimates $p_u^{(k-2)}, p_u^{(k-1)}, p_u^{(k)}$ and the corresponding
function values $\calA^{(k-2)}, \calA^{(k-1)}, \calA^{(k)}$, the next
estimate is
\begin{equation}
  p_u^{(k+1)} = p_u^{(k)} - \frac{2c}{b \pm \sqrt{b^2 - 4ac}}   ,
\end{equation}
where $a, b, c$ are the coefficients of the quadratic interpolant
through the three points. The sign in the denominator is chosen to
maximize $|b \pm \sqrt{b^2 - 4ac}|$.
This method is well-suited for QNM finding because: (i)~it works in
the complex plane without requiring the Jacobian (unlike Newton's
method for $\mathbb{C}$-valued functions); (ii)~it converges
superlinearly (order $\approx 1.84$); and (iii)~it naturally handles
the near-degenerate cases where $|\calA|$ has a flat minimum.
Convergence is declared when $|\calA(p_u)| < 10^{-10}$ or when
$|p_u^{(k+1)} - p_u^{(k)}| < 10^{-12}$.

\label{app:Richardson}
For the Richardson extrapolation, the frequency $\omega(z_\mathrm{max})$
is computed at four geometrically spaced values:
\begin{equation}
  z_\mathrm{max}^{(j)} = z_0 \times r^j   , \quad j = 0, 1, 2, 3   ,
\end{equation}
with $z_0$ and $r$ chosen so that $z_\mathrm{max}^{(0)} \approx 50\pi$
and $z_\mathrm{max}^{(3)} \approx 300\pi$ (dimension-dependent).
The extrapolation uses the ansatz
\begin{equation}
  \omega(z_\mathrm{max}) = \omega_\infty + c   z_\mathrm{max}^{-\alpha}
  + O(z_\mathrm{max}^{-2\alpha})   ,
\end{equation}
with $\alpha = 2d/(d+2)$. Each pair of data points
$(z_\mathrm{max}^{(i)}, z_\mathrm{max}^{(j)})$ gives an independent
estimate of $\omega_\infty$ via
\begin{equation}
  \omega_\infty^{(ij)} = \frac{\omega_i   (z_j/z_0)^\alpha
  - \omega_j   (z_i/z_0)^\alpha}{(z_j/z_0)^\alpha - (z_i/z_0)^\alpha}   .
\end{equation}
The final estimate is the average of all $\binom{4}{2} = 6$
pair-extrapolations, and the error is the standard deviation.

\label{app:stability}
Finally, the stability scan searches for QNMs with $\im(p_u) > 0$
(corresponding to $\im(\omega) > 0$, i.e., growing modes). The scan
evaluates $|\calA(p_u)|$ on a $15 \times 15$ grid in
$\re(p_u) \in (-5, 8)$, $\im(p_u) \in (0.01, 5.0)$;
grid points with $|\calA(p_u)| < 1.0$ trigger Muller
refinement, and only modes with $\im(p_u) > 0.001$ after
refinement are accepted (to exclude artifacts from Muller drifting
to the lower half-plane).
For $d = 3, 4, 5$, no accepted modes were found, confirming the absence
of unstable QNMs.